\documentclass[format=acmsmall, review=false, natbib=true, screen=true]{acmart}

\usepackage[utf8]{inputenc}
\usepackage{booktabs} 
\usepackage{tabularx}
\usepackage{subcaption}
\usepackage{hyperref}
\usepackage{pgfplots}
\usepackage{pgfplotstable}
\usepackage{datatool}
\usepackage{xfor}
\usepackage{siunitx} 
\usepackage{xstring}
\usepackage{float}
\usepackage{natbib}

\usepackage[ruled]{algorithm2e} 

\usepackage{xcolor}
\usepackage{verbatim}

\SetAlFnt{\small}
\SetAlCapFnt{\small}
\SetAlCapNameFnt{\small}
\SetAlCapHSkip{0pt}
\IncMargin{-\parindent}

\setcopyright{acmlicensed}

\renewcommand\footnotetextcopyrightpermission[1]{} 
\usepackage{fancyhdr}
\newcommand{\gru}{\textsc{gru4rec}\xspace}
\newcommand{\knn}{\textsc{s-knn}\xspace}

\newcommand{\sr}{\textsc{sr}\xspace}
\newcommand{\ar}{\textsc{ar}\xspace}
\newcommand{\mc}{\textsc{mc}\xspace}
\newcommand{\iknn}{\textsc{iknn}\xspace}
\newcommand{\sknn}{\textsc{sknn}\xspace}
\newcommand{\vsknn}{\textsc{v-sknn}\xspace}
\newcommand{\vknn}{\textsc{v-sknn}\xspace}

\newcommand{\ssknn}{\textsc{s-sknn}\xspace}
\newcommand{\sfknn}{\textsc{sf-sknn}\xspace}
\newcommand{\sfsknn}{\textsc{sf-sknn}\xspace}
\newcommand{\smf}{\textsc{smf}\xspace}
\newcommand{\fpmc}{\textsc{fpmc}\xspace}
\newcommand{\bpr}{\textsc{bpr-mf}\xspace}
\newcommand{\fism}{\textsc{fism}\xspace}
\newcommand{\fossil}{\textsc{fossil}\xspace}
\newcommand{\eq}{\textsc{eq}\xspace}

\definecolor{color1}{RGB}{43,131,186}
\definecolor{color2}{RGB}{171,221,164}
\definecolor{color3}{RGB}{215,25,28}
\definecolor{color4}{RGB}{253,174,97}

\newcommand{\td}[1]{%
	\IfSubStr{#1}{S}{%
		*\StrBehind{#1}{S}[\tmp]\StrLeft{\tmp}{6}[\tmpp]\num[math-rm=\mathbf]{\tmpp}%
	}{%
		\IfSubStr{#1}{F}{%
			\StrBehind{#1}{F}[\tmp]\StrLeft{\tmp}{6}[\tmpp]\num[math-rm=\mathbf]{\tmpp}%
		}{%
			\StrLeft{#1}{6}[\tmp]\num{\tmp}%
		}%
	}%
}%

\newcommand{\tdz}[1]{%
	\sisetup{%
		round-mode          	= places, 
		round-precision     	= 0, 
		detect-weight			= true,
		detect-inline-weight	= math
	}%
	\IfSubStr{#1}{S}{%
		*\StrBehind{#1}{S}[\tmp]\StrLeft{\tmp}{6}[\tmpp]\num[math-rm=\mathbf]{\tmpp}%
	}{%
		\IfSubStr{#1}{F}{%
			\StrBehind{#1}{F}[\tmp]\StrLeft{\tmp}{6}[\tmpp]\num[math-rm=\mathbf]{\tmpp}%
		}{%
			\StrLeft{#1}{6}[\tmp]\num{\tmp}%
		}%
	}%
}%
\newcommand{\tdn}[1]{%
	\sisetup{%
		round-mode          	= places, 
		round-precision     	= 2, 
		detect-weight			= true,
		detect-inline-weight	= math
	}%
	\IfSubStr{#1}{S}{%
		*\StrBehind{#1}{S}[\tmp]\StrLeft{\tmp}{6}[\tmpp]\num[math-rm=\mathbf]{\tmpp}%
	}{%
		\IfSubStr{#1}{F}{%
			\StrBehind{#1}{F}[\tmp]\StrLeft{\tmp}{6}[\tmpp]\num[math-rm=\mathbf]{\tmpp}%
		}{%
			\StrLeft{#1}{6}[\tmp]\num{\tmp}%
		}%
	}%
}%

\begin{document}

\title{Evaluation of Session-based Recommendation Algorithms}

\author{Malte Ludewig}
\affiliation{%
	\institution{TU Dortmund}
	\streetaddress{Otto-Hahn-Straße 12}
	\city{Dortmund}
	\state{44227}
	\country{Germany}
}
\email{malte.ludewig@tu-dortmund.de}

\author{Dietmar Jannach}
\affiliation{%
	\institution{AAU Klagenfurt}
	\streetaddress{Universitätsstraße 65-67}
	\city{Klagenfurt am Wörthersee}
	\state{9020}
	\country{Austria}
}
\email{dietmar.jannach@aau.at}

\date{}

\begin{abstract}
Recommender systems help users find relevant items of interest, for example on e-commerce or media streaming sites. Most academic research is concerned with approaches that personalize the recommendations according to long-term user profiles. In many real-world applications, however, such long-term profiles often do not exist and recommendations therefore have to be made solely based on the observed behavior of a user during an ongoing session.
Given the high practical relevance of the problem, an increased interest in this problem can be observed in recent years, leading to a number of proposals for \emph{session-based recommendation algorithms} that typically aim to predict the user's immediate next actions.

In this work, we present the results of an in-depth performance comparison of a number of such algorithms, using a variety of datasets and evaluation measures. Our comparison includes the most recent approaches based on recurrent neural networks like \gru, factorized Markov model approaches such as \fism or \fossil, as well as 
simpler methods based, e.g., on nearest neighbor schemes. Our experiments reveal that algorithms of this latter class, despite their sometimes almost trivial nature, often perform equally well or significantly better than today's more complex approaches based on deep neural networks. Our results therefore suggest that there is substantial room for improvement regarding the development of more sophisticated session-based recommendation algorithms.\footnote{A preliminary comparison of sequential recommendation algorithms was presented in our own previous work in \citep{JannachLudewig2017RecSys,KamehkhoshJannachLudewigRecTemp2017}.}
\end{abstract}

\ccsdesc[500]{Information systems~Recommender systems}
\ccsdesc[500]{General and reference~Evaluation}

\keywords{Session-based Recommendation; Sequential Recommendation; Deep Learning; Factorized Markov Models, Nearest Neighbors}

\maketitle

\section{Introduction}
\label{sec:introduction}

Many of today's online services use recommender systems to point their users or site visitors to additional items that might be of interest to them. In academic research, the majority of works is focusing on techniques that rely on long-term preference models to determine the items to be presented to the user. However, in many application domains of recommender systems, such long-term user models are often not available for a larger fraction of the users, e.g., because they are first-time visitors or because they are not logged in. Consequently, suitable recommendations have to be determined based on other types of information, usually the user's most recent interactions with the site or application. Recommendation techniques that rely solely on the user's actions in an ongoing session and which adapt their recommendations to the user's actions are called \emph{session-based} recommendation approaches \citep{QuadranaetalCSUR2018}.

Amazon's ``Customers who bought \dots~also bought'' recommendations can be considered an extreme case of such a session-based approach. In this case, the recommendations are seemingly only dependent on the item that is currently viewed by the user (and the purchasing patterns of the community). A number of other techniques were proposed in the research literature, which do not limit themselves to the very last action, but consider some or all user actions since the session started.
Some of these techniques only consider which events happened; others, in contrast, in addition take the sequence of events into account in their algorithms. Besides the e-commerce domain, a number of other application fields were in the focus in the literature, among them in particular music, web page navigation, or travel and tourism.

In academia, sequential recommendation problems are typically operationalized as the task of predicting the next user action.
Experimental evaluations are usually based on larger, time-ordered logs of user actions, e.g., on the users' item viewing and purchase activities on an e-commerce shop or on their listening history on a music streaming site. From an algorithmic perspective, early approaches to predict the next user actions were based, for example, on sequential pattern mining techniques. Later on, different types of more sophisticated methods based on Markov models were proposed and successfully applied to the problem. Finally, in the most recent years, the use of deep learning approaches based on artificial neural networks was explored as another solution. Recurrent Neural Networks (RNN), which are capable of learning models from sequentially ordered data, are a ``natural choice'' for this problem, and significant advances regarding the prediction accuracy of such algorithms were reported in the recent literature \citep{Hidasi2015,Tan2016,Hidasi2016,Hidasi2017,Devooght2017}.

Despite the growing number of papers on the topic in recent years, no true ``standard'' benchmark data sets or evaluation protocols exist in the community. Therefore, it remains difficult to compare the various algorithmic proposals, in particular as often different baseline algorithms are used in the papers. And, for some of them it is also unclear if they are particularly strong. In our previous work \citep{JannachLudewig2017RecSys,KamehkhoshJannachLudewigRecTemp2017}, we could, for example, demonstrate that a comparably simple k-nearest-neighbor method leads to similar or even better accuracy results than a modern deep learning approach.

To establish a common base for future research, we performed an in-depth performance comparison across multiple domains and datasets, which involved a number of comparably simple as well as more sophisticated algorithms from the recent literature. Our results show that computationally and conceptually simple methods often lead to predictions that are similarly accurate or even better than those of today's most recent techniques based on deep learning models. As a consequence, we argue that researchers should take these simpler methods as alternative baselines into account when developing novel session-based recommendation algorithms. Furthermore, our results suggest that there is still substantial room for improvement regarding the development of more sophisticated session-based recommendation algorithms.

This paper extends our previous works presented in \citep{JannachLudewig2017RecSys,KamehkhoshJannachLudewigRecTemp2017} in a number of ways. First, we made experiments for a larger number of datasets from different domains, using a richer set of performance measures. Second, we included recent \emph{sequential} recommendation algorithms like FISM and FOSSIL \citep{kabbur13fism,he16fusing} in the evaluation as well as the latest version of \gru \citep{Hidasi2015}. Third, we designed a number of additional sequence-aware similarity measures for the previously proposed session-based nearest neighbor method, which in most cases lead to significant performance gains. Finally, we also propose a new method called \emph{Session-based Matrix Factorization (SFM)}, which yields good results in some of the tested application domains.

The paper is organized as follows. Next, in Section \ref{sec:related-work}, we discuss previous works and typical application areas of session-based recommendation approaches. In Section \ref{sec:methods}, we provide technical details about the algorithms that were compared in our work. Section \ref{sec:evaluation-setup} describes our evaluation setup and Section \ref{sec:results} the outcomes of our experiments. To foster reproducible research on the topic, we share the code of the used evaluation framework and the compared algorithms online.\footnote{\url{https://www.dropbox.com/sh/dbzmtq4zhzbj5o9/AACldzQWbw-igKjcPTBI6ZPAa?dl=0}}
\newpage

\section{Review of Session-Based Recommendation Approaches}
\label{sec:related-work}
Most of the approaches for session-based recommendation proposed in the literature implement some form of \emph{sequence learning}, see also \citep{QuadranaetalCSUR2018} for a recent survey on the more general class of sequence-aware recommenders. Early approaches were based on the identification of \emph{frequent sequential patterns}, which can be used at recommendation time to predict a user's next action. These early approaches were applied, for example, in the context of predicting the online navigation behavior of users \citep{Mobasher2002}. Later on, such pattern mining techniques were also used for next-item recommendation problems in e-commerce or the music domain \citep{yap12sequential,hariri12context,Bonnin2014}.

While frequent pattern techniques are easy to implement and lead to interpretable models, the mining process can be computationally demanding. At the same time, finding good algorithm parameters, in particular a suitable minimum support threshold, can be challenging.
Finally, in some application domains it seems that using frequent item sequences does not lead to better recommendations than when using simpler item co-occurrence patterns \citep{Bonnin2014}. In the context of this work, we investigate both sequential 
and co-occurrence patterns in their simplest forms as baselines.

In many newer works, more sophisticated sequence learning approaches were proposed that implement some form of \emph{sequence modeling}. Such sequence modeling approaches are usually based on Markov Chain (MC) models \citep{he09query,mcfee11nlp,garcin13news,hosseinzadeh15adapting}, reinforcement learning (RL) and Markov Decision Processes (MDP) \citep{shani05mdp,moling12optimal,tavakol14fmdp}, or Recurrent Neural Networks (RNN) \citep{zhang13sequential,sordoni15hierarchical,Hidasi2015,Hidasi2016,liu16unified,song16multirate,twardowski16contextual,yu16dynamic,du16recurrent,soh17deep}.
Again, the typical application scenarios of these methods include the e-commerce and the music recommendation domain.

An early approach based on an MDP model was proposed by \cite{shani05mdp}. It demonstrated the value of using sequential data in an e-commerce scenario, but also showed that models based on Markov Chains often cannot be directly applied due to data sparsity. Therefore, \cite{shani05mdp} proposed different heuristics to overcome the problem. An additional challenge when using this type of models 
is to decide how many preceding interactions should be considered when predicting the next one. Some authors therefore use a mixture of Variable-order Markov Models (VMMs) or \emph{context-trees} to consider sequences of different lengths \citep{he09query,garcin13news}. Other works, for example by \cite{hosseinzadeh15adapting}, rely on Hidden Markov Models (HMMs) to overcome certain limitations of plain Markov Chain models. In \citep{shani05mdp,moling12optimal}, reinforcement learning was implemented based on MDPs, which made it possible to also consider the reward for the shop in the recommendation process. To deal with the problem of the explosion of the state space in such scenarios, \cite{tavakol14fmdp} proposed to model the state space based on the sequence of item attributes in order to predict the characteristics of the next item that the user will consider.  In the context of the comparative analysis presented in this paper, we limit ourselves to a simple MC-based method as a baseline, in particular because some techniques like the one discussed by \cite{tavakol14fmdp} require the existence of knowledge about certain item attributes.

The most recent works on sequence modeling are based on RNNs. \cite{zhang13sequential}, for example, used them for the prediction of user clicks in an advertisement scenario. \cite{Hidasi2015} were among the first to explore Gated Recurrent Units (GRUs) as a special form of RNNs for the prediction of the next user action in a session. Their method called \gru was later on extended in different ways in \citep{Hidasi2016,Hidasi2017} and \citep{quadrana17personalizing}. While \cite{Hidasi2015} reported substantial performance improvements over an \emph{item-based} k-nearest-neighbor (kNN) method when using their first version of \gru, our previous work \citep{JannachLudewig2017RecSys} showed that a \emph{session-based} nearest neighbor method also leads to competitive accuracy results for the same problem setting. Since \gru was substantially improved since its initial version, we include the latest version of the method proposed by \cite{Hidasi2017} in the performance comparison reported in this paper.
Furthermore, given our observations regarding the often competitive performance of conceptually simpler methods we designed a number of variations of the basic session-based nearest neighborhood method from \citep{JannachLudewig2017RecSys}, which we also considered in the experiments.

Another family of sequence modeling approaches relies on \emph{distributed item representations}, e.g., in the form of latent Markov embeddings \citep{chen12embedding,chen13multispace,wu13karaoke,feng15personalized} or distributional embeddings
\citep{djuric14hidden,baezayates15next,grbovic15prod2vec,tagami15modeling,vasile16meta,reddy16learning,zheleva2010www}. Embeddings are dense, lower-dimensional representations that are derived from sequentially ordered data and encode transition probabilities based on the observations in the original data. They were applied, for example, in the domains of next-track music recommendation \citep{zheleva2010www,chen12embedding}, recommendation of learning courses \citep{reddy16learning}, or next point-of-interest (POI) recommendation \citep{feng15personalized}.
However, a general challenge when using item embeddings is that they can be computationally demanding and sometimes require substantial amounts of training data to be effective. In the context of our work, we experimented with item embeddings as an alternative representation of the user sessions. However, the usage of embeddings did not lead to an improvement in terms of the prediction accuracy for our problem settings, which is why we do not report the detailed outcomes of these experiments in this paper.

To overcome the limitations of pure sequence learning methods, a number of \emph{hybrid methods} were proposed that, for instance, combine the advantages of matrix factorization techniques with sequence modeling approaches in the form of Factorized Markov Chains \citep{rendle10FPMC,lian13checkin,cheng13where,he16inferring,he16fusing}. \cite{rendle10FPMC} proposed the Factorized Personalized Markov Chain (\fpmc) approach as an early method for next-item recommendations in e-commerce settings, where user interactions are represented as a three-dimensional tensor (user, current item, next-item). Later on, variations of \fpmc were proposed and successfully applied for a variety of application problems, e.g., by \cite{kabbur13fism} and \cite{he16fusing}. Other hybrid techniques that, for example, use some form of clustering or Latent Dirichlet Allocation in combination with a sequential recommendation method were proposed, e.g., in \citep{hariri12context,natarajan13app,song15next}, for the problems of next-track or next-app recommendation. In our experimental evaluation, we include both the \fpmc method by \cite{rendle10FPMC} as well as the recent variations and improvements described by \cite{kabbur13fism} (\fism) and \cite{he16fusing} (\fossil).

Besides pure \emph{session-based} techniques, which solely consider a user's action of the ongoing session, there are also approaches that consider previous interactions of the same user in the recommendation process. Such techniques are called \emph{session-aware} according to the terminology of \cite{QuadranaetalCSUR2018}. Examples of such works include \citep{baezayates15next,Billsus:2000:LAW:325737.325768,hariri12context,Jannach2015,JannachKamehkhoshEtAl2017,quadrana17personalizing}, and session-aware approaches were applied for various application domains like e-commerce, music, news, or next-app recommendation. Considering longer-term user preferences in these papers shows to be helpful to improve the recommendations in the current, ongoing session. In some cases, like in \citep{Jannach2015}, it however turns out that the short-term user intents are much more important than the longer-term models. In the research presented herein, we therefore exclusively focus on session-based recommendation scenarios. We however consider the combination of long-term and short-term models as an important area for future research.

\section{Details of the Investigated Methods}
\label{sec:methods}
Based on these discussion, we include the following four types of techniques in our comparison of session-based recommendation algorithms: simple heuristics as baseline methods, nearest-neighbor techniques, recurrent neural networks, and 
factorization-based methods. The main input to all methods is a training set of past user sessions, where each session consists of a set of sequentially ordered actions of a given type, e.g., an item view event in an online shop or a consumption event on a media streaming site. The models learned by the algorithms can then be used to predict the next event in a given user session in the test set. In our evaluations, we follow a pragmatic approach to determine user sessions---in case these are not provided in the datasets---and use user inactivity times to determine session borders. The details for each dataset are described later in this paper.

Regarding the choice of the algorithms, we focus on collaborative filtering methods based on implicit feedback signals, e.g., item view or music listening events. Depending on the specific application, content-based and hybrid algorithms can be designed that use additional meta-data or content features. Since these features are domain specific and such features are only available for very few of our datasets, we limit ourselves to methods that do not rely on such types of data in this paper.

\subsection{Baseline Methods}
\label{sec:baselines}
We include the following baseline techniques in our comparison: a method that we call Simple Association Rules (\ar), first-order Markov Chains (\mc), and a method that we named Sequential Rules (\sr). All baselines implement very simple prediction schemes, have a low computational complexity both for training and recommending, and only consider the very last item of a current user session to make the predictions. 
Furthermore, we include a prediction method based on Bayesian Personalized Ranking (\bpr) proposed by \cite{Rendle2009} as an alternative baseline.

\subsubsection{Simple Association Rules (\ar)}
\label{baselines:ar}
Simple Association Rules (\ar) are a simplified version of the association rule mining technique \citep{Agrawal:1993:MAR:170035.170072} with a maximum rule size of two. The method is designed to capture the frequency of two co-occurring events, e.g., ``Customers who bought~\dots~also bought''. Algorithmically, the rules and their corresponding importance are ``learned'' by counting how often the items $i$ and $j$ occurred together in a session of any user.

Let a session $s$ be a chronologically ordered tuple of item click events $s=(s_1,s_2,s_3,\dots,s_m)$ and $S_p$ the set of all past sessions.
Given a user's current session $s$ with $s_{|s|}$ being the last item interaction in $s$, we can define the score for a recommendable item $i$ as follows, where the indicator function $1_{\eq}(a,b)$ is $1$ in case $a$ and $b$ refer to the same item and $0$ otherwise.
\small
\begin{equation}
score_{\ar}(i,s) = \frac{1}{\sum_{p \in S_p} \sum_{x=1}^{\vert p\vert} 1_{\eq}(s_{\vert s\vert},p_{x}) \cdot (|p|-1)} \sum_{p \in S_p} \sum_{x=1}^{\vert p\vert} \sum_{y=1}^{\vert p\vert} 1_{\eq}(s_{\vert s\vert},p_x) \cdot 1_{\eq}(i,p_y)
\label{eq:ar}
\end{equation}
\normalsize
In Equation \ref{eq:ar}, the sums at the right-hand side represent the counting scheme. The term at the left-hand side normalizes the score by the number of total rule occurrences originating from the current item $s_{|s|}$.
A list of recommendations returned by the \ar method then contains the items with the highest scores in descending order. No minimum support or confidence thresholds are applied.
In our implementation, as shared online, we create the rules in one iteration over the training data and store them (sorted by weight) in nested maps to support fast lookups in the recommendation phase. With this data structure, top-n recommendations can be created almost instantaneously.

\subsubsection{Markov Chains (\mc)}
The \mc baseline can be seen as a variant of \ar with a focus on sequences in the data. Here, the rules are extracted from a first-order Markov Chain, see \citep{markovchainnorris97}, which describes the transition probability between two \emph{subsequent} events in a session. In our baseline approach, we simply count how often users viewed item $q$ immediately after viewing item $p$.
Technically, the score for an item $i$ given the current session $s$ with the last event $s_{|s|}$ can be defined as a simplified version of Equation \ref{eq:ar}:
\small
\begin{equation}
score_{\mc}(i,s) = \frac{1}{\sum_{p \in S_p} \sum_{x=1}^{\vert p\vert-1} 1_{\eq}(s_{\vert s\vert},p_{x})} \sum_{p \in S_p} \sum_{x=1}^{\vert p\vert-1} 1_{\eq}(s_{\vert s\vert},p_{x}) \cdot 1_{\eq}(i,p_{x+1})
\label{eq:mc}
\end{equation}
\normalsize
where the function $1_{\eq}(a,b)$ again indicates whether $a$ and $b$ refer to the same item or not.
Here, with the right-hand side of the formula, we count how often item $i$ appears immediately after $s_{|s|}$. The normalization term transforms the absolute count into a relative transition probability.
In line with \ar, in our implementation the rules and weights are recorded in nested maps in one single iteration over the training data to ensure short training times and to support the fast generation of the recommendations.

\subsubsection{Sequential Rules (\sr)}
\label{sec:sequential_rules}
Finally, the \sr method as proposed in \citep{KamehkhoshJannachLudewigRecTemp2017} is a variation of \mc or \ar respectively.
It also takes the order of actions into account, but in a less restrictive manner.
In contrast to the \mc method, we create a rule when an item $q$ appeared after an item $p$ in a session even when other events happened between $p$ and $q$.

When assigning weights to the rules, we consider the number of elements appearing between $p$ and $q$ in the session. Specifically, we use the weight function $\textstyle w_{\sr}(x) = 1/(x)$, where $x$ corresponds to the number of steps between the two items.\footnote{Other weighting functions, e.g., with a logarithmic decay, are possible as well. Using the linear function however led to the best results, on average, in our experiments.} Given the current session $s$, the \sr method calculates the score for the target item $i$ as follows:
\small
\begin{equation}
score_{\sr}(i,s) = \frac{1}{\sum_{p \in S_p} \sum_{x=2}^{\vert p\vert} 1_{\eq}(s_{\vert s\vert},p_{x}) \cdot x} \sum_{p \in S_p} \sum_{x=2}^{\vert p\vert} \sum_{y=1}^{x-1} 1_{\eq}(s_{\vert s\vert},p_y) \cdot 1_{\eq}(i,p_x) \cdot w_{\sr}(x-y)
\label{eq:sr}
\end{equation}
\normalsize
In contrast to Equation \ref{eq:ar} for \ar, the third inner sum only considers indices of previous item view events for each session $p$. In addition, the weighting function $w_{\sr}(x)$ is added. Again, we normalize the absolute score by the total number of rule occurrences for the current item $s_{|s|}$. As for \ar and \mc, the algorithm was implemented using nested sorted maps, which can be created in a single iteration over the training data.

\subsubsection{Bayesian Personalized Ranking (\bpr)}
\label{baselines:bpr}
To make our results comparable with previous research, we finally include a prediction method based on \bpr as a baseline in our experiments.\footnote{The method was proposed by Hidasi et al.~in the context of the \gru method.}
\bpr proposed by \cite{Rendle2009} is a learning-to-rank method designed for implicit-feedback recommendation scenarios. The method is usually applied for matrix-completion problem formulations based on longer-term user-item interactions.
In \bpr the matrix is factorized into two smaller matrices of latent user and item features ($W$ and $H$),  optimizing the following criterion:
\begin{equation}
	BPR_{OPT} = \sum_{(u,i,j) \in D_S} ln \, \sigma( r_{u,i} - r_{u,j} ) - \lambda_{\Theta}||\Theta||^{2}
	\label{eq:bpt_opt}
\end{equation}
In the above formula, a ranking $r_{u,i}$ for user $u$ and item $i$ is approximated with the dot product of the corresponding rows in the matrices $W$ and $H$ ($ r_{u,i} = \langle W_u, H_i \rangle $).
The model parameters $\Theta = (W,H)$ are learned using stochastic gradient descent in multiple iterations over the dataset $D_S$, which consists of triplets of the form $(u,i,j)$, where $(u,i)$ is a positive feedback pair and $(u,j)$ is a sampled negative example. The optimization criterion in Equation \ref{eq:bpt_opt} aims to rank the positive sample $(u,i)$ higher than a non-observed sample $(u,j)$.

To apply the method for the session-based recommendation scenario---where there are no long-term user profiles---we attribute each session in the training set to a different user, i.e., each session corresponds to a user in the user-item interaction matrix. At prediction time, we use the average of the latent item vectors of the current session so far as the user vector.

Generally, BPR and other methods designed for the matrix-completion problems in their original form, i.e., without considering the short-term session context, do not lead to competitive results in session-based recommendation scenarios, as reported, e.g., in \citep{Jannach2015}. Therefore, we do not consider such algorithms, e.g., traditional matrix factorization techniques, as baselines in our experiments.

\subsection{Nearest Neighbors}
Despite their simplicity, nearest-neighbor-based approaches often perform surprisingly well as discussed, e.g., by \cite{Verstrepen:2014:UNN:2645710.2645731} and in our previous work \citep{JannachLudewig2017RecSys,KamehkhoshJannachLudewigRecTemp2017}. We, therefore, include different nearest neighbor schemes in our comparison. First, we consider a more traditional item-based variant, which was also employed as a baseline method by \cite{Hidasi2015}. Furthermore, we evaluate three variations of a more recent session-based nearest neighbor technique in our experiments.

\subsubsection{Item-based kNN (\iknn)}

The \iknn method as used in \citep{Hidasi2015} only considers the last element in a given session and then returns those items as recommendations that are most similar to it in terms of their co-occurrence in other sessions. Technically, each item is encoded as a binary vector, where each element corresponds to a session and is set to ``1'' in case the item appeared in the session. The similarity of two items can then be determined, e.g., using the cosine similarity measure, and the number of neighbours $k$ is implicitly defined by the desired recommendation list length.

Conceptually, the method implements a certain form of a ``Customers who bought~\dots~also bought'' scheme like the \ar baseline. The use of the cosine similarity metric however makes it less susceptible to popularity biases. Although item-to-item approaches are comparably simple, they are commonly used in practice and sometimes considered a strong baselines \citep{Linden2003, Davidson2010}.
In terms of the technical implementation, all similarity values can be pre-computed and sorted in the training process to ensure fast responses at recommendation time.\footnote{We use the implementation published at \url{https://github.com/hidasib/GRU4Rec}.}

\subsubsection{Session-based kNN (\sknn)}
\label{sec:sknn}
Instead of considering only the last event in the current session, the \sknn method compares the entire current session with the past sessions in the training data to determine the items to be recommended, see also \citep{hariri12context,Bonnin2014,Lerche2016}. Technically, given a session $s$, we first determine the $k$ most similar past sessions (neighbors) $N_s$ by applying a suitable session similarity measure, e.g., the Jaccard index or cosine similarity on binary vectors over the item space \citep{Bonnin2014}. In our experiments, the binary cosine similarity measure led to the best results. As in \citep{JannachLudewig2017RecSys}, using $k=500$ as the number of neighbors to consider led to good performance results for many datasets.
Next, given the current session $s$, its neighbors $N_s$, and the chosen similarity function $sim(s_1,s_2)$ for two sessions $s_1$ and $s_2$, the recommendation score for each item $i$ can as defined by \cite{Bonnin2014}:
\begin{equation}
score_{\sknn}(i,s) = \Sigma_{n \in N_s} sim(s,n) \cdot 1_{n}(i) 
\label{eq:knn}
\end{equation}
Here, the indicator function $1_{n}(i)$ returns $1$ if session $n$ contains item $i$ and $0$ otherwise.

\paragraph{Scalability Considerations.}
Given a current session $s$, we cannot scan a potentially large set of past sessions for possible neighbors in an online recommendation scenario. Therefore, in our implementation of the algorithm, as described in \citep{JannachLudewig2017RecSys} in more detail, we rely on pre-computed in-memory index data structures and on neighborhood sampling to enable fast recommendation responses. The index is used to quickly locate past sessions that contain a certain item, i.e., the index allows us to retrieve \emph{possible} neighbor sessions that contain at least one element of the current session through fast lookup operations. On the other hand, sampling only a smaller fraction of all past sessions in our experiments as potential neighbors has shown to lead to comparably small accuracy compromises. In fact, in some domains like e-commerce, only looking for neighbors in the most recent sessions---thereby capturing recent trends in the community---proved to be very effective \citep{JannachLudewigLerche2017umuai} and led to even better results than when all past sessions were taken into account.

Our nearest neighbor implementations, therefore, have an additional parameter $m$, which determines the size of the sample from which the neighbors of a target session are taken. In the experiments reported in \citep{JannachLudewig2017RecSys}, it was, for example, sufficient to consider only the 1,000 most recent sessions from several million existing ones.

\paragraph{Sequence-Aware Extensions: \vsknn, \ssknn, and \sfknn}
The described \sknn method does not consider the order of the elements in a session when using the Jaccard index or cosine similarity as a distance measure. Since the order of the elements might, however, be relevant in some domains and since the user preferences might change within a single session depending on the already seen items, we propose three variations of the \sknn method.\footnote{We made additional experiments using other ways of encoding sequential information, e.g., by using embeddings of sessions and items with the popular \emph{Word2Vec} and \emph{Doc2Vec} approaches. However, none of these variations led to better accuracy results than the \sknn method in our experiments. We therefore omit these results from our later discussions.}

\begin{itemize}
\item Vector Multiplication Session-Based kNN (\vsknn): The idea of this variant is to put more emphasis on the more recent events of a session when computing the similarities. Instead of encoding a session as a \emph{binary} vector as described above, we use real-valued vectors to encode the current session. Only the very last element of the session obtains a value of ``1''; the weights of the other elements are determined using a linear decay function that depends on the position of the element within the session, where elements appearing earlier in the session obtain a lower weight. As a result, when using the \emph{dot product} as a similarity function between the current weight-encoded session and a binary-encoded past session, more emphasis is given to elements that appear later in the sessions.
\item Sequential Session-based kNN (\ssknn): This variant also puts more weight on elements that appear later in the session. This time, however, we achieve the effect with the following scoring function:
\begin{equation}
score_{\ssknn}(i,s) = \Sigma_{n \in N_s} sim(s,n) \cdot w_{n}(s) \cdot 1_{n}(i)
\label{eq:ssknn}
\end{equation}
Here, the indicator function $1_{n}(i)$ is complemented with a weighting function $w_{n}(i,s)$, which takes the order of the events in the current session $s$ into account. The weight $w_{n}(i,s)$ increases when the more recent items of the current session $s$ also appeared in a neighboring session $n$. If an item $s_{x}$ is the most recent item of the current session $s$ that also appears in the neighbor session $n$, then the weight will be defined as $w_{n}(s) = x / |s| $,
where the index $x$ indicates the position of $s_{x}$ within the session.\footnote{Note that the weighting function is designed to work independently from the similarity function. We rely on the binary session representation for the similarity calculation without considering the order of the items to ensure computational efficiency.} If, for example, the second-to-last item of the current session with a length of $5$ is the most recent item also included in the neighbor session $n$, the weight would be $w_{n}(i,s)=4/5$. Items from this neighbor can, therefore, potentially obtain a higher score than, e.g., items from neighbor sessions that only include the third from last item of the current session, which are assigned a weight of $3/5$.
\item Sequential Filter Session-based kNN (\sfknn):  This method also uses a modified scoring function, but in a more restrictive way. The basic idea is that given the last event (and related item $s_{|s|}$) of the current session $s$, we only consider items for recommendation that appeared directly after $s_{|s|}$ in the training data at least once.
\begin{equation}
score_{\sfknn}(i,s) = \Sigma_{n \in N_s} sim(s,n) \cdot 1_{n}(s_{\vert s\vert},i)
\label{eq:sfknn}
\end{equation}
While the general scoring function is identical to the one of \sknn (Equation \ref{eq:knn}), we use a different implementation of the indicator function $1_{n}(s_{|s|},i)$. Here, $1$ is only returned if there exists any past session which contains the sequence $(s_{|s|},i)$, given $s_{|s|}$ is the item currently viewed in the user's current session $s$. Though the sequence $(s_{|s|},i)$ can be part of any past session, the item $i$ obviously still has to be a part of the neighbor session $n$ for the indicator function to return $1$.
\end{itemize}

\subsection{Neural Networks -- \gru}
\label{sec:neural_net}

Approaches based on Recurrent Neural Networks (RNNs), as discussed in Section \ref{sec:related-work}, represent the most recently explored family of techniques for session-based recommendation problems. Among these methods, \gru is one of the latest deep learning approaches that was specifically designed for session-based recommendation scenarios \citep{Hidasi2015,Hidasi2017}.

\begin{figure}[t!]
	\centering
	\includegraphics[width=.7\textwidth]{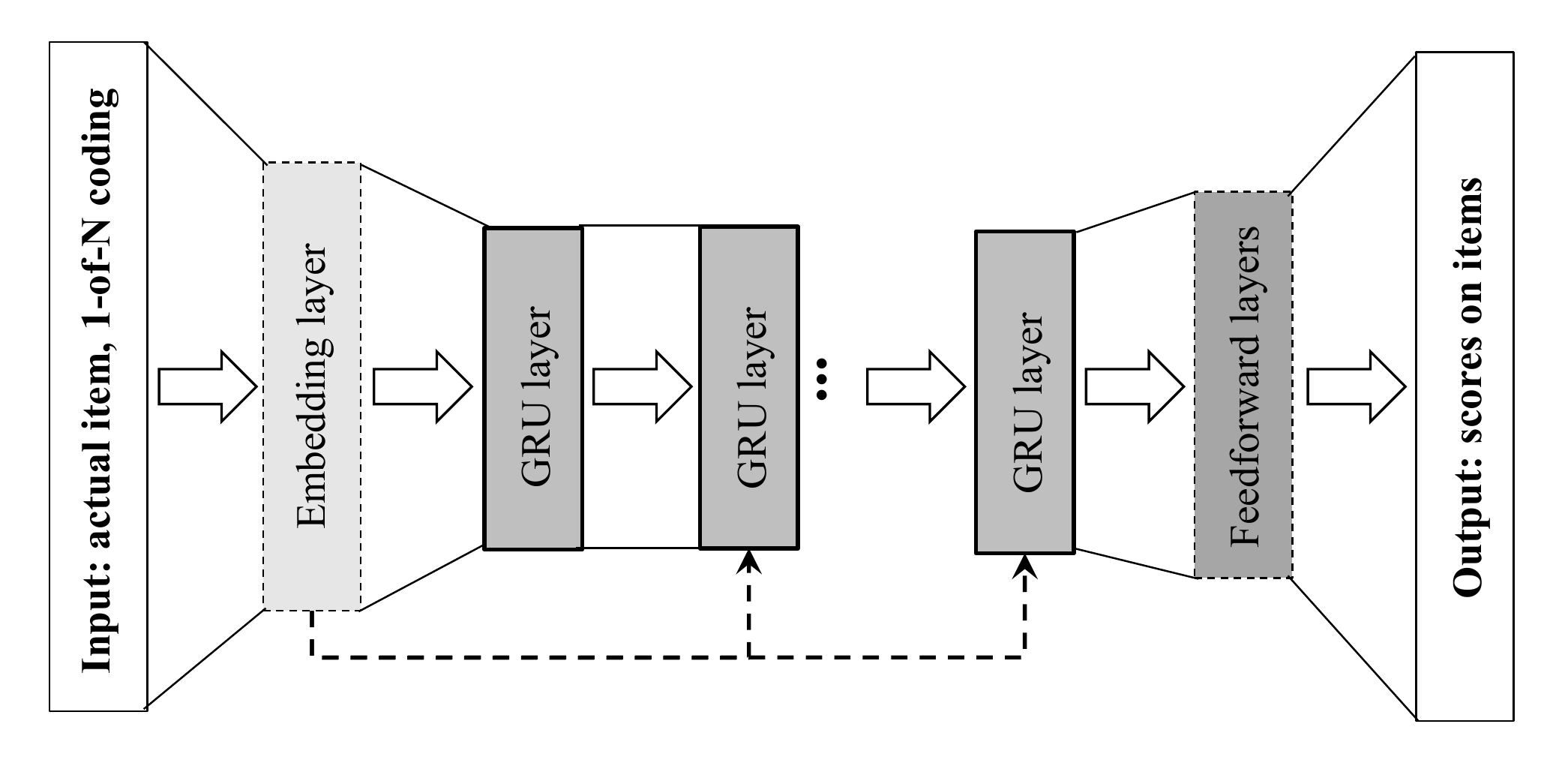} 
	\caption{Architecture of the \gru neural network, adapted from \citep{Hidasi2015}.}
	\label{fig:gru_architecture}
\end{figure}

\gru models user sessions with the help of an RNN with Gated Recurrent Units \citep{cho2014gru} in order to predict the probability of the subsequent events (e.g., item clicks) given a session beginning. Figure \ref{fig:gru_architecture} shows the general architecture of the network, in which the embedding, the feedforward, and additional GRU layers are optional. In fact, the authors of the method found that a single GRU layer of varying width led to the best performance in their experiments.

The input of the network is formed by a single item, which is one-hot encoded in a vector representing the entire item space, and the output is a vector of similar shape that should give a ranking distribution for the subsequent item. Inbetween, the standard GRU layer keeps track of a hidden state that encodes the previously occurring items in the same session. Therefore, while training and predicting with the help of this network architecture, the items of a session have to be fed into the network in the correct order and the hidden state of the GRUs has to be reset after a session ends. In terms of the activation functions, the authors found $tanh$ and the $sigmoid$ function to work best for the GRU and the ranking layer, respectively.

While the usage of RNNs for session-based, or more generally, sequential prediction problems is a natural choice, the particular network architecture, the choice of the loss functions, and the use of session-parallel mini-batches to speed up the training phase are key innovative elements of the approach.

\begin{figure}[t!]
	\centering
	\includegraphics[width=.85\textwidth]{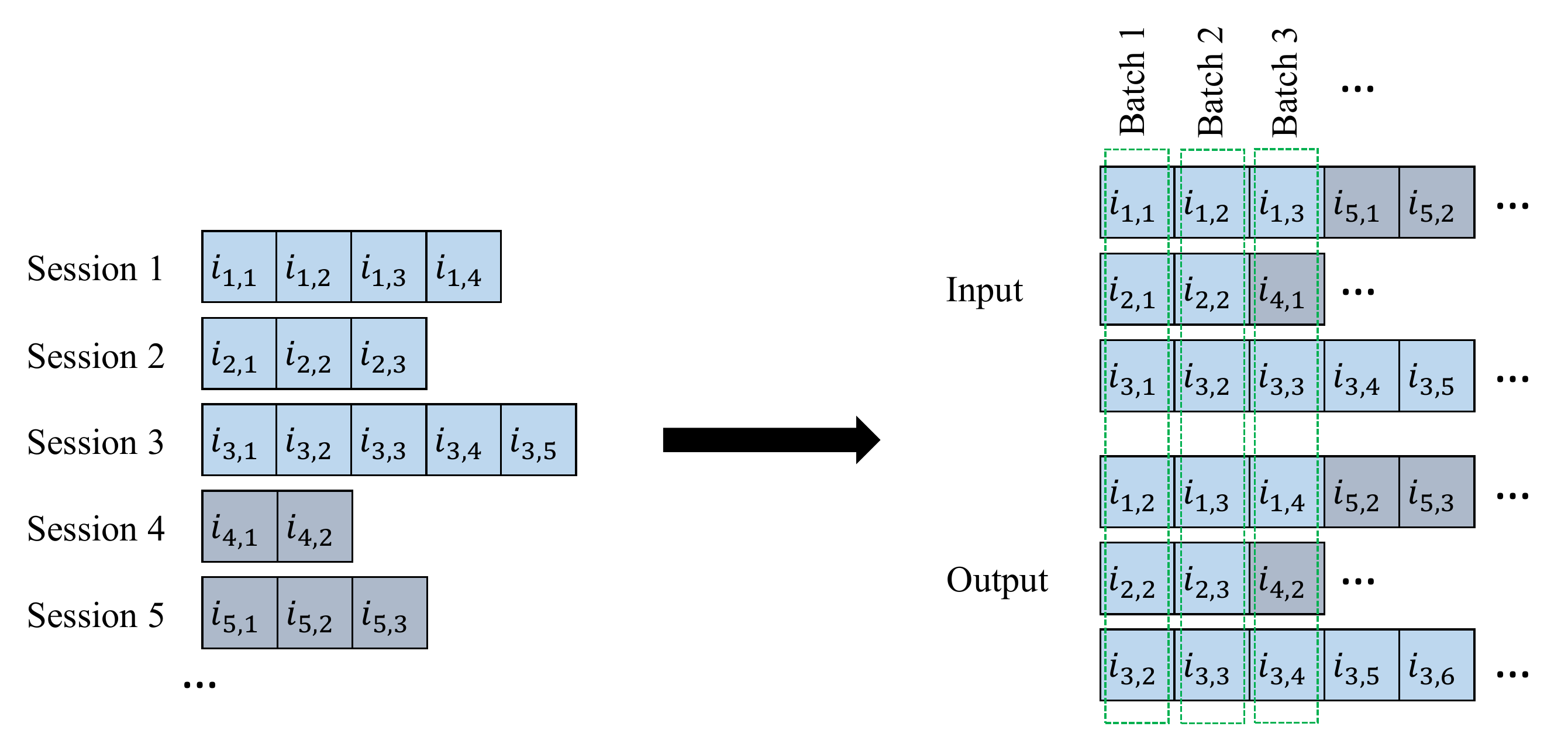} 
	\caption{Illustration of the session-parallel mini-batch scheme of \gru, adapted from \citep{Hidasi2015}.}
	\label{fig:gru_sessionbatch}
\end{figure}

The model can be trained with stochastic gradient descent (SGD) using established optimizations like \emph{ADAM}, \emph{ADADELTA}, \emph{RMSProp}, or \emph{ADAGRAD} \citep{Duchi2011adagrad,Zeiler2012adadelta,Kingma2014adam}. As common in practice when optimizing deep neural networks, Hidasi et al. train the network in batches. To ensure that the items or events are fed into the network in the correct order, they propose the \emph{session-parallel mini-batch} training scheme, which is illustrated in Figure \ref{fig:gru_sessionbatch}. 
In the training process, each part of a batch belongs to a specific session in the training data and the network records a separate hidden state for each position.
Whenever a session at a position in the batch ends, the corresponding hidden state is reset and the next batch update includes the first event of a new session at that position.

As usual, a number of hyper-parameters can be tuned, including, the learning rate, the layer sizes, a momentum factor, and a drop-out factor to stabilize the network. The choice of the loss function is another key to the quality of the recommendations of \gru. The following loss functions were designed or applied by the authors. In particular the latest function (\emph{MAX}) proposed by \cite{Hidasi2017} led to a significant performance improvement over the previous ones.

\begin{itemize}
	\item \emph{BPR}: Bayesian Personalized Ranking (BPR), as discussed above, uses a pairwise ranking loss function for the task of creating top-n recommendations. In \gru, a generalized version of this function is applied using the following formula:
	\begin{equation}
		L_s(\hat{r}_{s,i},S_N) = -\frac{1}{|S_N|} \cdot \sum_{j \in S_N} log(\sigma(\hat{r}_{s,i}-\hat{r}_{s,j}))
	\end{equation}
	In the loss function, the predicted rating $\hat{r}_{s,i}$ for the actual next item $i$ given the current session $s$ is compared to a set of negative samples $S_N$ with the goal of maximizing the difference between them. Here, the sigmoid and logarithm functions are applied to represent the proportion between the ranking of the negative and the positive example.
	\item \emph{TOP1}: This loss function was introduced by the authors of \gru and can be seen as a regularized approximation of the relative rank of a positive sample $\hat{r}_{s,i}$ and the negative samples $S_N$:
	\begin{equation}
	L_s(\hat{r}_{s,i},S_N) = \frac{1}{|S_N|} \cdot \sum_{j \in S_N} \sigma(\hat{r}_{s,j}-\hat{r}_{s,i}) + \sigma(\hat{r}_{s,j}^2)
	\end{equation}
	Here, the proportion is approximated with the sigmoid function, and the regularization term $\sigma(\hat{r}_{s,j}^2)$ is added so that the score of the negative samples is directed to zero.
	\item \emph{MAX}: In continuation of their work, the authors proposed a generic extension to these two loss functions, where $L_s$ stands for a loss function like \emph{BPR} or \emph{TOP1} defined above:
	\begin{equation}
	L_{max}(\hat{r}_{s,i},S_N) = L_s( \hat{r}_{s,i},\{\max_{j \in S_N} \hat{r}_{s,j}\} )
	\end{equation}
	Instead of using a sum of differences between the positive item's rating $\hat{r}_{s,i}$ and the negative samples $S_N$, only the highest rated negative sample $\max_{j \in S_N} \hat{r}_{s,j}$ from $S_N$ is used to calculate the loss. As this function has to be differentiable for SGD training, $\max_{j \in S_N}$ is approximated with the \emph{softmax} function. The resulting functions $BPR_{max}$ and $TOP1_{max}$ showed superior performance when compared to the \emph{BPR} and \emph{TOP1} functions \citep{Hidasi2017}.
\end{itemize}

In our experiments, we used the \gru (v2.0) implementation that the authors shared online.
The code is regularly maintained by the authors and includes the implementation of the \gru method, the code of their baseline algorithms, as well as the code for the evaluation procedure proposed in \citep{Hidasi2015}.

\subsection{Factorization-based Methods}
As described in Section \ref{sec:related-work}, a number of (hybrid) factorization-based methods were proposed in recent years for \emph{sequential} recommendation problems. We include three existing methods from the literature in our experiments, Factorized Personalized Markov Chains (\fpmc) proposed by \cite{rendle10FPMC}, \fism by \cite{kabbur13fism}, and \fossil by \cite{he16fusing}. Generally, these methods aim at predicting the next actions of users, but were not designed for session-based recommendation scenarios with anonymous users. We therefore describe for each method how we applied them to our problem setting. In addition, we propose a novel factorization method called Session-based Matrix Factorization (\smf), which relies on the $BPR_{max}$ and $TOP1_{max}$ loss functions as described above.

\subsubsection{Factorized Personalized Markov Chains (\fpmc)}
The \fpmc method was designed for the specific problem of next-basket recommendation. The problem consists of predicting the contents of the next basket of a user, given his or her history of past shopping baskets. By limiting the basket size to one item and by considering the current session as the history of baskets, the method can be directly applied for session-based recommendation problems.

\begin{figure}[H]
	\centering
	\includegraphics[width=.65\textwidth]{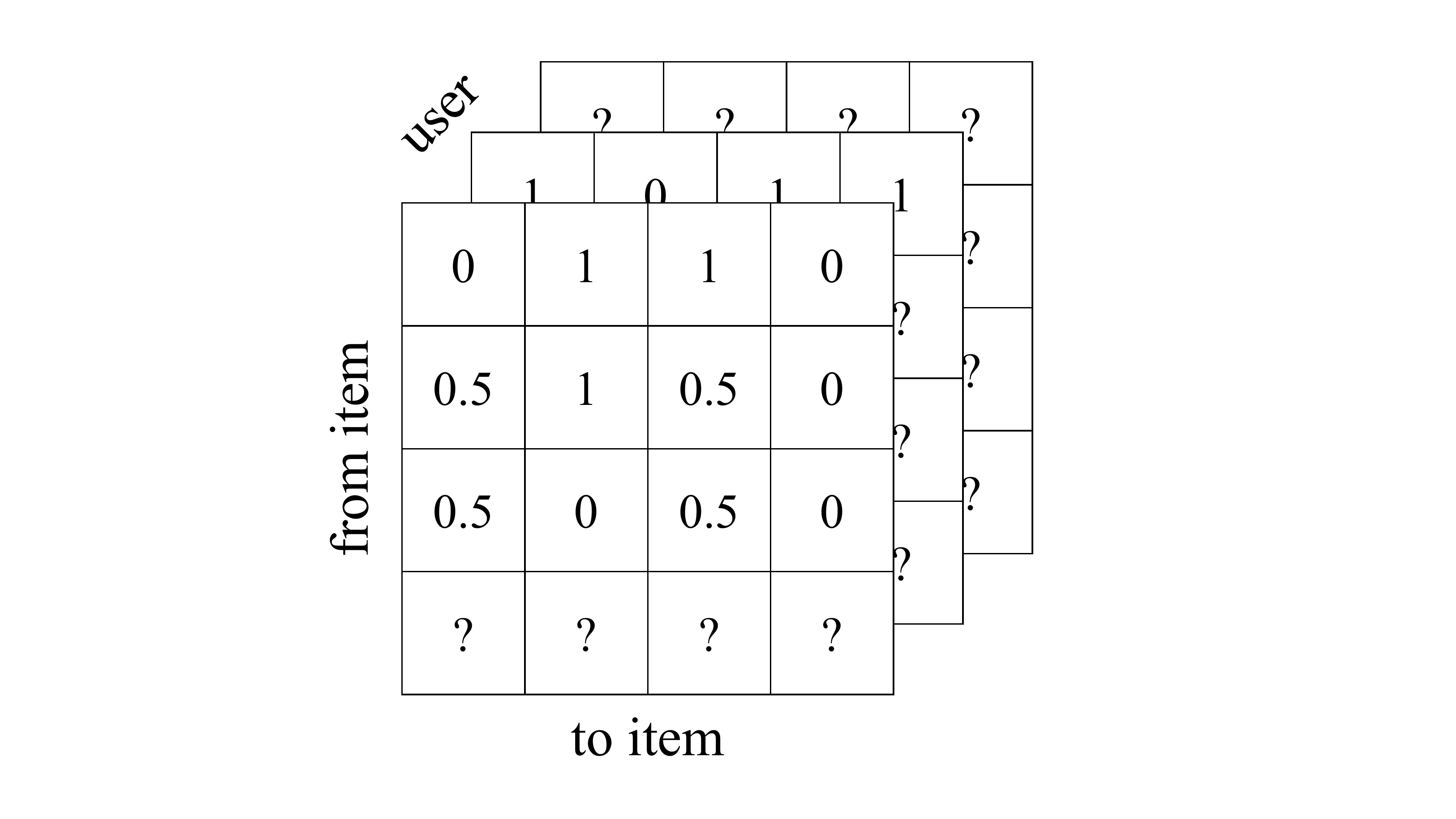}
	\captionof{figure}{Personalized transition cube, adapted from \citep{Rendle2009}.}
	\label{fig:fpmc_cube}
\end{figure}

Technically, \fpmc combines \mc and traditional user-item matrix factorization in a three dimensional tensor factorization approach. As illustrated in Figure \ref{fig:fpmc_cube}, the third dimension captures the transition probabilities from one item to another.

Internally, a special form of the Canonical Tensor Decomposition is used to factor the cube into latent matrices, which can then be used to predict a ranking in the following way:
\begin{equation}
\hat{r}_{u,l,i} = \langle v_{u}^{U,I},v_{i}^{I,U}\rangle + \langle v_{i}^{I,L},v_{l}^{L,I}\rangle + \langle v_{u}^{U,L},v_{l}^{L,U}\rangle
\end{equation}
where $\hat{r}_{u,l,i}$ is a score for item $i$ with the preferences of user $u$ when he or she previously examined item $l$.
The three-dimensional decomposition results in six latent matrices $v^{X,Y}$ representing the latent factors for dimension $X$ regarding dimension $Y$, e.g., $v^{U,L}$ are the user latent factors in terms of the previously examined item and $v^{I,L}$ the item latent factors regarding the previously examined item. Accordingly, $v_{u}^{U,L}$ for example represents the factors for a single user $u$ and $v_{i}^{I,L}$ the factors for item $i$, which are combined with the regular dot product ($\langle a,b \rangle$) to calculate the ranking $\hat{r}_{u,l,i}$.
Those latent factors are learned using SGD with the pairwise ranking loss function BPR.

In our problem setting, where we have no long-term user histories, each session in the training data corresponds to a user. Once the model is trained, each new session therefore represents a user cold-start situation. To apply the model to our setting, we estimate the session latent vectors as the average of the latent factors of the individual items in the session. This approach was adopted also by \cite{Hidasi2015} to apply \bpr to session-based recommendation scenarios.

\subsubsection{Factored Item Similarity Models (\fism)}
This method is based on an item-item factorization, which has the advantage of being
directly applicable to our session-based cold-start scenario, where no explicit
user representation can be  learned. However, \fism does not incorporate
sequential item-to-item transitions like \fpmc does. Equation \ref{eq:fism} shows the prediction function which \cite{kabbur13fism} trained using SGD to predict ratings, e.g., for the movie domain.
\begin{equation}
\hat{r}_{u,i} = b_u + b_i + (n_u^+)^{-\alpha} \sum_{j \in R_u^+} p_j q_i^T
\label{eq:fism}
\end{equation}
Technically, for user $u$ and item $i$, a score $\hat{r}_{u,i} $ is calculated as the sum of latent vector products $p_j q_i^T$ between item $i$ and the items $R_u^+$ already rated by the user $u$. In our scenario, $R_u^+$ corresponds to the previously inspected items in a session. The terms $b_u$ and $b_i$ are bias terms and $n_u^+$ specifies the number of ratings by user $u$, which is combined with a parameter $\alpha$ to normalize the sum of vector products to a certain degree.
Instead of using the $RMSE$ as an error metric, we use $BPR$'s pairwise loss function when optimizing the top-n recommendations for the given implicit feedback scenario.

\subsubsection{Factorized Sequential Prediction with Item Similarity Models (\fossil)}
In this approach, \fism is combined with factorized Markov chains to incorporate sequential information into the model.
The model can be described as shown in Equation \ref{eq:fossil} (from \cite{he16fusing}):
\begin{equation}
\hat{r}_{u,l,i} = \underbrace{ \sum_{j \in R_u^+ \setminus \{i\}} p_j q_i^T }_\text{long-term preferences} + \overbrace{(w + w_u)}^\text{personalized weighting} \cdot \underbrace{n_l m_i^T}_\text{sequential dynamics}
\label{eq:fossil}
\end{equation}
Again, $\hat{r}_{u,l,i} $ represents a rating for item $i$ given a user $u$ and his or her previously inspected item $l$. The first term represents the long-term user preferences and corresponds to the \fism model in Equation \ref{eq:fism}. Using a weighted sum with a global factor $w$ and a personalized factor $w_u$, the model is extended by a factorized Markov chain to capture the sequential dynamics. In the last term of Equation \ref{eq:fossil}, a latent vector $n_l$ for item $l$ is multiplied with a latent vector $m_i$ for item $i$ to factor in the user-independent probability of item $l$ being followed by item $i$.

In our scenario, again, the sessions represent the users, $R_u$ corresponds to the current session and $BPR$ is used as the loss function to rank suitable items over negative examples.

\subsubsection{Session-based Matrix Factorization (\smf)}

Finally, \smf is a novel factorization-based model that we designed for the specific task of session-based recommendation. Similar to \fossil it combines factorized Markov chains with classic matrix factorization. In addition, our method considers the cold-start situation of session-based recommendation scenarios as follows.

In contrast to the traditional factorization-based prediction model $r_{u,i} = p_u q_i^T$, in the \smf method, we replace the latent user vector $p_u$ with a session preference vector $s_e$, which is computed as an embedding of the current session $s$:
\begin{equation}
s_e = M_{ST} \cdot s^T
\end{equation}
Here, the session $s$ is as a binary vector similar to the representation in \sknn (see Section \ref{sec:sknn}) and $M_{ST}$ is a transformation matrix of size $|I| \cdot |u_s|$, which reduces the size of the binary session vector (number of unique items $|I|$) to a specific latent vector size $|s_e|$.

Based on the embedded session representation $s_e$, the prediction function is defined as shown in Equation \ref{eq:smf}.
\begin{equation}
\hat{r}_{s,l,i} = w_i\cdot( \underbrace{ s_e q_i^T + b_{1,i}}_\text{session preferences} ) + (1-w_i)\cdot( \overbrace{n_l m_i^T + b_{2,i}}^\text{sequential dynamics})
\label{eq:smf}
\end{equation}

The score $\hat{r}_{s,l,i}$ for a session $s$ with the most recent item $l$ and an item $i$ is computed as a weighted combination of session preferences and sequential dynamics.
Here, the session preferences correspond to the long-term user preferences in the traditional matrix factorization model, i.e., the embedded session latent vector $s_e$ for the current session $s$ is multiplied with an item latent vector $q_i$ for item $i$ to compute a relevance score $i$ regarding $s$. The sequential dynamics are captured exactly as in Equation \ref{eq:fossil} for \fossil using latent representations for the currently inspected item $l$ and item $i$. Both partial scores are adjusted with a separate bias term  $b_{x,i}$ and combined in a weighted sum with the factor $w_i$ dependent on item $i$.

To train this model, we incorporated some of the concepts from \gru (see Section \ref{sec:neural_net}). Specifically, we adopted \emph{ADAGRAD} for SGD-based optimization, and used $BPR_{max}$ and $TOP1_{max}$ as loss functions.
Furthermore, we integrated two additional concepts (and corresponding hyper-parameters) in the training phase to avoid model over-fitting: a session drop-out factor and a skip-rate. For a drop-out factor of $0.1$, for example, each positive entry of the binary session input vector is set to 0 with a probability of 10\%.
The skip-rate, in contrast, describes how often not the immediate next item in the log data should be used as a positive sample in the training process, but the subsequent one. A skip rate of $0.1$ therefore means that in 10\,\% of the cases the immediate next item is skipped.


\section{Experiment Setup}
\label{sec:evaluation-setup}
In this section, we describe the details of our algorithm comparison in terms of to the used evaluation protocol, the performance measures, and the evaluation datasets. All source code and pointers to the public datasets are provided online to ensure reproducibility of our research.\footnote{\url{https://www.dropbox.com/sh/dbzmtq4zhzbj5o9/AACldzQWbw-igKjcPTBI6ZPAa?dl=0}}

\subsection{Evaluation Protocol and Performance Measures}
\label{sec:protocol}
The general computational task in session-based recommendation problems is to generate a ranked list of objects that in some form ``matches'' a given session beginning. What represents a good match, depends on the specific application scenario. It could be a set of alternative shopping items in an e-commerce scenario or a continuation of given music listening session.

In offline evaluations for session-based recommendations, researchers often abstract from the underlying purpose of the system \citep{JannachAdomavicius2016}, e.g., if the recommender should help discover something new or find alternatives to a currently inspected item.
Instead, the recorded user sessions are typically considered as a ``gold standard'' for the evaluation. To measure the performance of an algorithm, researchers resort to assessing the capability of an algorithm to predict the withheld entries of a session.

Different approaches are found in the literature to withhold certain entries of a session. In some works, only the last element is hidden \citep{hariri12context,Bonnin2014}, some propose to ``reveal'' the first $n$ elements of a session \citep{Jannach2015}, while others, finally, evaluate their approaches by iteratively revealing one entry 
after the other \citep{Hidasi2016}. We employed the latter iterative revealing scheme in our experiments as it (i) conceptually includes both of the other techniques and (ii) reflects the user journey throughout a session in the best way.

\paragraph{Selection of the Target Item and Accuracy Measures.}
We measured prediction accuracy in two ways and correspondingly report the results in separate tables.
\begin{itemize}
\item First, to establish comparability with existing research, we use an evaluation scheme in which the task is to predict the \emph{immediate next item} given the first $n$ elements. For each session, we iteratively increment $n$, measure the hit rate (HR) and the Mean Reciprocal Rank (MRR), and finally determine the average HR and MRR for all sessions for the different list lengths, as done by \cite{Hidasi2016}.
\item Second, instead of focusing only on the next item, we made a measurement where we considered \emph{all subsequent} elements for the given session beginning, because all of them might be relevant to the user. In this scheme, we used the standard information retrieval measures precision and recall at defined list lengths. The number of given elements of the session is also iteratively incremented as in the previously described evaluation scheme.
\end{itemize}

\paragraph{Sessionization strategies.} Different strategies exist in the literature to split the user activity logs into sessions. In some of the public datasets used in our evaluation, the activity logs were already split up into sessions, i.e., each log entry was assigned a unique session ID (\emph{RSC15}, \emph{Zalando}). For other datasets (\emph{RETAILR}, \emph{NOWPLAYING}, \emph{30MUSIC}, \emph{CLEF}), we used a common heuristic-based approach and considered a session as over after a defined user idle time, e.g., 30 minutes of user inactivity \citep{Cooley1999}.
For the TMALL dataset, where the timetamps for the recorded events were only available at the granularity of a day, we considered all events of one day as belonging to one session. Finally, for the two playlists dataset (\emph{AOTM}, \emph{8TRACKS}), we considered all elements of a playlist to be part of a session.

\paragraph{Training and Test Splits, Repeated Subsampling.} \cite{Hidasi2016} used one single training-test split. In the case of an e-commerce dataset, the data was split in a way that the sessions of all six months except those of the very last day of the entire dataset were placed in the training set. The last day was used for testing.
We report the results of applying this evaluation scheme to ensure comparability, e.g., with respect to the results obtained for the e-commerce dataset that was used in their experiments.

Since such single-split setups have their limitations, we focus our discussion on the results that were obtained when applying a \emph{sliding-window} protocol, where we split the data into 5 slices of equal size in days. For most e-commerce data, for example, we  used the data of about one month for training and the subsequent data (e.g., of one day) for testing (see Section \ref{sec:datasets} for the dataset specific configurations). This allows us to make multiple measurements with different test sets. We then evaluate the performance for each of these data samples and report the average of the performance results for all slices. This latter protocol helps us reduce the danger that the observed outcomes are the results of one particular train-test configuration.\footnote{To ensure that the smaller size of those splits does not negatively affect the performance of the model-based approaches, we tested the single-split configurations as well on all datasets. The obtained results are mostly in line with those obtained with the sliding-window protocol and shown in Appendix \ref{appx:singleres}.}

For the playlist datasets \emph{8TRACKS} and \emph{AOTM} no timestamp information is available. For these datasets we therefore applied a standard cross-validation procedure, where elements are randomly assigned to the training and test sets. We did not use such a time-agnostic data splitting procedure for the e-commerce and news datasets for different reasons. First, as the results will show, there are strong temporal effects that should be considered in the recommendation process. Second, in these domains, the set of items is not static and in particular in the news domain new items appear constantly. Randomly splitting the sessions would then potentially result in the effect that future interactions with not-yet-existing items would be considered in the training phase.

\paragraph{Additional Quality Factors.} Since accuracy is not the only relevant quality factor in practice, we made the following additional measurements, as was done by \cite{JannachLudewig2017RecSys}.

\begin{itemize}
\item \emph{Coverage:} We report how many different items ever appear in the top-$k$ recommendations. This measure represents a form of catalog coverage, which is sometimes referred to as \emph{aggregate diversity} \citep{Adomavicius:2012:IAR:2197072.2197127}.
\item \emph{Popularity bias:} High accuracy values can, depending on the measurement method, correlate with the tendency of an algorithm to recommend mostly popular items \citep{JannachLercheEtAl2015}.
To assess the popularity tendencies of the tested algorithms, we report the \emph{average popularity score} for the elements of the top-$k$ recommendations of each algorithm. 
This average score is the mean of the \emph{individual popularity scores} of each recommended item. We compute these scores based on the training set by counting how often each item appears in one of the training sessions and by then applying min-max normalization to obtain a score between 0 and 1.


\item \emph{Cold start:} Some methods might only be effective when a significant amount of training data is available. We, therefore, report the results of measurements where we artificially removed parts of the (older) training data to simulate such situations.
\item \emph{Scalability:} Training modern machine learning methods can be computationally challenging, and obtaining good results may furthermore require extensive parameter tuning. We, therefore, report the times that the algorithms needed to train the models and to make predictions at runtime. In addition, we report the memory requirements of the algorithms.
\end{itemize}

By reporting quality factors \emph{coverage} and \emph{popularity bias} our aim is to emphasize that different recommendation strategies can lead to quite different recommendations, even if they are similar in terms of the prediction accuracy, see also \citep{JannachLercheEtAl2015}. Such multi-metric evaluation approaches should also help practitioners to better understand the potential side effects of the recommenders, e.g., reduced average sales diversity and additionally increased sales of top-sellers \citep{DBLP:conf/icis/LeeH14}. It remains however difficult to aggregate the individual performance factors into one single score, as the relative importance of the factors can depend not only on the application domain, but also on the specific business model of the provider.

\paragraph{Parameter Optimization.}
Some of the algorithms that we tested require extensive (hyper-)parameter tuning including \smf and \gru. Thus, we systematically optimized the parameters for those algorithms for each dataset. Due to the computational complexity of the methods, we restricted the layer size for \gru as well as the number of latent factors for \smf to 100 and used a randomized search method with 100 iterations for the remaining parameters as described by \cite{Hidasi2017}. In each iteration, the learning rate, the drop-out factor, the momentum, and the loss function were determined in a randomized process to find the maximum hit rate for a list length of 20. All optimizations were performed on special validation splits, which were created by splitting a  training set into a validation training and test set.
For the simpler \knn-based approaches, we used the same validation sets to manually adjust the number of neighbors and samples when applying cosine similarity as the distance measure (except for \vknn).
The final parameters for each method and dataset are provided in Appendix \ref{appx:params}.

\subsection{Datasets}
\label{sec:datasets}
We made measurements for datasets from three different domains: e-commerce, music, and news.

\noindent \paragraph{E-Commerce Datasets.} We used the following four e-commerce datasets.
\begin{itemize}
  \item \emph{RSC15.} This is one of the datasets that was used in \citep{Hidasi2015} and their later works. It was published in the context of the ACM RecSys 2015 Challenge and contains recorded click sequences (item views, purchases) for a period of six months. We use the label \emph{RSC15-S} to denote the dataset and measurement where only one single train-test split is used. For RSC15, each split consists of 30 days of training and 1 day of test data.
  \item \emph{TMALL.} This dataset was published in the context of the TMall competition and contains interaction logs of the tmall.com website for one year. For TMALL, each split consists of 90 days of training and 1 day of test data.
  \item \emph{RETAILR.} The e-commerce personalization company \emph{retailrocket} published this dataset covering six month of user browsing activities, also in the context of a competition. For RETAILR, each split consists of 25 days of training and 2 days of test data.
  \item \emph{ZALANDO.} The final dataset is non-public and was shared with us by the fashion retailer Zalando. It contains user logs of their shopping platform for a period of one year. In our evaluation, we only considered the item \emph{view} events as was done for the other e-commerce datasets. For ZALANDO, each split consists of 90 days of training and 1 day of test data.
\end{itemize}

\begin{table}[!t]
	\centering
	\footnotesize
	\caption{Characteristics of the e-commerce datasets. The values are averaged over all five non-overlapping splits for each dataset, except for \emph{RSC15-S}, where we only use one train-test split.}
	\label{tab:charac_ecommerce}
	\begin{tabular}{@{}lrrrrr@{}}
		\toprule
		Dataset            & RSC15-S & RSC15       & TMALL  & RETAILR  & ZALANDO     \\ \midrule
		Actions            & 31.71M     & 5.43M     & 13.42M    & 212,182       & 4.54M    \\
		Sessions           & 7.98M      & 1.38M    & 1.77M   & 59,962      & 365,126      \\
		Items              & 37,483        & 28,582     & 425,348    & 31,968        & 189,328    \\
		Timespan in Days   & 182          & 31          & 91          & 27           & 91          \\
		\midrule
		Actions per Session & 3.97   & 3.95 & 7.56 & 3.54  & 12.43 \\
		Unique Items per Session  & 3.17  & 3.17 & 5.56 & 2.56  & 8.39 \\
		Actions per Day     & 174,222  & 175,063 & 149,096 & 7,858  & 50,410 \\
		Sessions per Day   & 43,854  & 44,358 & 19,719 & 2,220  & 4056 \\
		\bottomrule
	\end{tabular}
\end{table}

Table \ref{tab:charac_ecommerce} shows an overview of the characteristics of the e-commerce datasets. Except for the \emph{RSC15-S} dataset, which we include to make our evaluation comparable with previous works \citep{Hidasi2017,JannachLudewig2017RecSys}, we report the average values after creating five data splits as described above.

\noindent \paragraph{Media Datasets: Music and News.}
As in \citep{JannachLudewig2017RecSys}, we use the music domain as an alternative
area to evaluate session-based recommendation algorithms, because music is commonly consumed within listening sessions in sequential order. We use the same datasets that were used in \citep{JannachLudewig2017RecSys}, which consist of two sets of \emph{listening logs} and two datasets of user-created \emph{playlists}. In addition, we made measurements using a dataset from the news recommendation domain.

We in particular consider the news domain because it has certain distinct characteristics \citep{KarimiIPM2018}. First, constantly new items become available for recommendation \citep{Das:2007:GNP:1242572.1242610,Liu:2010:PNR:1719970.1719976}. At the same time, items can also quickly become outdated. Second, previous research indicates that short-term popularity trends can be important for the success of a recommender \citep{DBLP:conf/clef/Ludmann17}. The experiments based on this dataset should therefore provide an indicator if the general insights obtained from other domains generalize to a domain with very specific characteristics.

\begin{itemize}
\item \emph{8TRACKS and AOTM:} These dataset include playlists created by music enthusiasts. The \emph{AOTM} dataset was collected from the Art-of-the-Mix platform and is publicly available \citep{DBLP:conf/ismir/McFeeL12}. The non-public \emph{8TRACKS)} dataset was shared with us by the 8tracks.com music platform. For all music datasets, each split consists of 90 days of training and 5 days of test data.

\item \emph{30MUSIC and NOWPLAYING:} The \emph{30MUSIC} dataset contains listening histories of the last.fm music platform and was published by \cite{DBLP:conf/recsys/TurrinQCPC15}. The \emph{NOWPLAYING} dataset was created from music-related tweets, where users posted which tracks they were currently listening \citep{NPDataset}.
\item \emph{CLEF:} The dataset was made available to participants of the 2017 CLEF NewsREEL challenge.\footnote{\url{http://www.clef-newsreel.org/}} It consists of a stream of user actions (e.g., article reads) and article publication events, which were collected by the company \emph{plista} for several publishers. In our evaluation we only considered the article read events. We used the data of the publisher with the largest amount of recorded interactions (the popular German sports news portal Sport1\footnote{\url{https://www.sport1.de/}}). For CLEF, each split consists of 5 days of training and 1 days of test data.

\end{itemize}

The statistics for the datasets from the media (music and news) domain are given in Table \ref{tab:charac_music}.

\begin{table}[h!t]
	\centering
    \footnotesize
	\caption{Characteristics of the music and news datasets. The values are again averaged over all five non-overlapping splits.}
	\label{tab:charac_music}
	\begin{tabular}{@{}lrrrrr@{}}
		\toprule
		& 8TRACKS     & 30MUSIC     & AOTM        & NOWPLAYING & CLEF     \\ \midrule
		Actions            & 1.50M   & 638,933      & 306,830    & 271,177 &  5.54M  \\
		Sessions           & 132,453    & 37,333       & 21,888     & 27,005  &   1.64M \\
		Items              & 376,422    & 210,633    & 91,166     & 75,169   &  742 \\
		Timespan in Days          & 95          & 95          & 95          & 95     &   6 \\
	 	\midrule
		Actions per Session & 11.32 & 17.11 & 14.02 & 10.04 & 3.37 \\
		Items per Session  & 11.31 & 14.47 & 14.01  & 9.38 & 3.17 \\
		Actions per Day     & 16,663 & 7,099 & 3,409 & 3,013 & 923,414 \\
		Sessions per Day   & 1,472 & 415 & 243 & 300 & 274,074 \\
		\bottomrule
	\end{tabular}
\end{table}

\section{Results}
\label{sec:results}

\subsection{E-Commerce Datasets}

Table \ref{tab:ecommerce-results} shows the MRR and Hit Rate results at a recommendation list length 20 for the four tested e-commerce datasets. In addition, we report the results when applying the standard measures precision and recall when considering \emph{all} hidden elements in the rest of the session as described above (see Table \ref{tab:results_ecommerce_prm}). Finally, we also report coverage and popularity statistics for each algorithm.

\sisetup{
	round-mode          	= places, 
	round-precision     	= 3, 
	detect-weight			= true,
	detect-inline-weight	= math
}
\begin{table}
	\caption{Hit rate (HR), Mean reciprocal rank (MRR), catalog coverage (COV), and the average popularity (POP) for a list length of 20 obtained for the e-commerce datasets. The table rows are ordered by MRR@20. The best values are highlighted in each column and, in case of accuracy measures, marked with a star when the difference w.r.t.~to the second-best performing method was statistically significant.}
	\setlength\tabcolsep{1.8pt}
	\label{tab:ecommerce-results}
	\begin{minipage}[t]{0.5\linewidth}
		\scriptsize
		\centering
		\subcaption{RSC15}
		\label{tab:rsc15_results}
		\DTLsetseparator{;}
		\begin{filecontents*}{tables_window_rsc15-window-best.csv.txt}
Metrics;HR@20;MRR@20;COV@20;POP@20
\gru;S0.6827005909116645;F0.30789428190205526;0.504039835618509;0.05411351067344
\sr;0.6530894167710559;0.3037278898069343;0.6676971869752596;0.07173445550669796
\smf;0.6658590500604008;0.30160555779283443;0.5652627166223958;0.05452361320392
\mc;0.6417675095977053;0.3004527747031581;0.6454013441313895;0.07006175707830245
\ar;0.6360975162362191;0.28940288998029523;0.6296337381245627;0.09263263229541731
\vsknn;0.653047697882291;0.28285040371599657;0.6186833129914595;0.07858419574508514
\ssknn;0.6019050987276808;0.271644120330393;0.6548861183222503;0.0715691476452437
\sfsknn;0.5890664415470067;0.27049846972652813;0.6185417120322706;0.06564053234057832
\knn;0.6213461557695853;0.26580908249900953;0.6341462486884317;0.07282213330356804
\iknn;0.4864531487845086;0.20815879016693958;0.7553519133191935;0.04085817645946
\fpmc;0.3626607679028523;0.20115953727771635;F0.9754395184353486;0.05452189890273999
\bpr;0.23470894008242116;0.17618355389083393;0.9107021355564786;0.08801180217766
\fism;0.16192690179983;0.11450143376232307;0.9740755956075656;F0.008265875804446002
\fossil;0.18966904034389861;0.0616403012899282;0.9167840088465716;0.04822120227172
\end{filecontents*}
\DTLloaddb[noheader]{rsc15}{tables_window_rsc15-window-best.csv.txt}
		\begin{tabular}{@{}lrrrr@{}}
			\toprule
			\DTLforeach
			{rsc15}
			{\key=Column1,\hit=Column2,\mrr=Column3,\cov=Column4,\pop=Column5}{%
				\ifthenelse{\value{DTLrowi}=1}{%
					\key & \mrr & \hit & \cov & \pop \\ \midrule
				}{%
					\key & \td{\mrr} & \td{\hit} & \td{\cov} & \td{\pop}
					\ifthenelse{\value{DTLrowi}=\DTLrowcount{rsc15}}{\\ \bottomrule }{\\ }%
				}%
			}%
		\end{tabular}
		\hspace{6pt}
	\end{minipage}
	\begin{minipage}[t]{0.49\linewidth}
		\scriptsize
		\centering
		\subcaption{TMALL}
		\label{tab:tmall_results}
		\DTLsetseparator{;}
		\begin{filecontents*}{tables_window_tmall-window-best.csv.txt}
Metrics;HR@20;MRR@20;COV@20;POP@20
\ssknn;0.386548310641567;S0.1852633664664242;0.4667524828908355;0.0245190514387
\knn;S0.403854261580295;0.18154697733677325;0.38079734188868175;0.02607249399888
\vsknn;0.37328817662165714;0.17891183986985457;0.4644749416593371;0.0235613578823
\bpr;0.2040177292981095;0.15906452430014956;0.722647056586653;0.05725515987402001
\sfsknn;0.21637739723950206;0.1357123365643923;0.4358756322494715;0.018155330220459998
\gru;0.2770651921317782;0.12905854416407286;0.1512617414449481;0.03547872534632
\ar;0.26210065936693716;0.1289027512900635;0.5091163154243468;0.02138930496486
\sr;0.2418561839271939;0.1276441463373797;0.5688123300036739;0.0214478367455
\smf;0.26059832014328316;0.12079287785807462;0.2611858959338655;0.0359308589187
\mc;0.20044819622393026;0.11563952769025705;0.49809714454809423;0.019378955072419997
\fpmc;0.11919433213343919;0.10074345528764599;F0.8803782985887505;0.005152245366364
\iknn;0.1500848625871168;0.05107961077242242;0.7280156301255649;0.007183148041346001
\fism;0.03677694918987455;0.023728188172821175;0.7522648894251058;F0.00262971992533
\fossil;0.003682974097943039;0.0014387370187966108;0.5981395097099222;0.015833349720514
\end{filecontents*}
\DTLloaddb[noheader]{tmall}{tables_window_tmall-window-best.csv.txt}
		\begin{tabular}{@{}lrrrr@{}}
			\toprule
			\DTLforeach
			{tmall}
			{\key=Column1,\hit=Column2,\mrr=Column3,\cov=Column4,\pop=Column5}{%
				\ifthenelse{\value{DTLrowi}=1}{%
					\key & \mrr & \hit & \cov & \pop \\ \midrule
				}{%
					\key & \td{\mrr} & \td{\hit} & \td{\cov} & \td{\pop}
					\ifthenelse{\value{DTLrowi}=\DTLrowcount{rsc15}}{\\ \bottomrule }{\\ }%
				}%
			}%
		\end{tabular}
	\end{minipage}
	\newline
	\vspace{3pt}
	\newline
	\begin{minipage}[t]{0.5\linewidth}
		\scriptsize
		\centering
		\subcaption{RETAILR}
		\label{tab:retailrockt_results}
		\DTLsetseparator{;}
		\begin{filecontents*}{tables_window_retailrocket-window-best.csv.txt}
Metrics;HR@20;MRR@20;COV@20;POP@20
\ssknn;S0.5906703739907334;S0.34506746724119275;0.5957839194075467;0.056183256891758994
\vsknn;0.5726725112410544;0.33788444217337327;0.5754070101128178;0.059879928733699996
\knn;0.5830103774790739;0.33712525618865125;0.5656942640210951;0.057722502292333576
\bpr;0.35699586753838963;0.3029696939204819;0.8241494678897267;0.0602334211569
\fpmc;0.3200561559658852;0.272939858441847;F0.9291334430165774;0.02246757220724
\sfsknn;0.3576308469641947;0.25996203947490326;0.40328434224636045;0.03454080303069104
\sr;0.41865502850410846;0.2454398209391674;0.5240526342607288;0.041944573734757205
\gru;0.47969637053202624;0.24279844402993164;0.6022763558055437;0.05978977404420001
\ar;0.4391178815682644;0.24117602121851003;0.5443568051078975;0.05272132075406975
\mc;0.358540255452433;0.2295765510797317;0.41064796270352505;0.0350279142368364
\smf;0.4593651727982039;0.22547437147844435;0.4490265039117016;0.08452739194884
\iknn;0.24014815722978833;0.10729263501616193;0.5844266045618054;0.033237416584179995
\fism;0.13168494963954674;0.07509563015903065;0.8482962095757642;F0.01764744558284
\fossil;0.05801168486827889;0.021528641132983046;0.7531080622423971;0.12698532909512
\end{filecontents*}
\DTLloaddb[noheader]{rrocket}{tables_window_retailrocket-window-best.csv.txt}
		\begin{tabular}{@{}lrrrr@{}}
			\toprule
			\DTLforeach
			{rrocket}
			{\key=Column1,\hit=Column2,\mrr=Column3,\cov=Column4,\pop=Column5}{%
				\ifthenelse{\value{DTLrowi}=1}{%
					\key & \mrr & \hit & \cov & \pop \\ \midrule
				}{%
					\key & \td{\mrr} & \td{\hit} & \td{\cov} & \td{\pop}
					\ifthenelse{\value{DTLrowi}=\DTLrowcount{rsc15}}{\\ \bottomrule }{\\ }%
				}%
			}%
		\end{tabular}
		\hspace{6pt}
	\end{minipage}
	\begin{minipage}[t]{0.49\linewidth}
		\scriptsize
		\centering
		\subcaption{ZALANDO}
		\label{tab:zalando_results}
		\DTLsetseparator{;}
		\begin{filecontents*}{tables_window_zalando-window-best.csv.txt}
Metrics;HR@20;MRR@20;COV@20;POP@20
\sr;0.4829648288337432;F0.30381945617847983;0.5860160610122639;0.06120370538597125
\mc;0.45538853591846673;0.3026460497714911;0.5128477968556637;0.05996548234936042
\iknn;0.4047707833461892;0.27458171552701216;0.7141048123093726;0.03694418450396
\gru;0.46786110038451234;0.2671595827595911;0.303881776173372;0.10060481126196001
\smf;0.4470251128044076;0.2668099507999785;0.36163483118542017;0.10721625983494001
\ar;0.4665121127562516;0.2578472438076795;0.46720725311093975;0.08863965468395987
\sfsknn;0.4377800918641637;0.24949445406725107;0.4324087313522864;0.05682229763714615
\vsknn;S0.5205166967210775;0.2326398968635311;0.4322725727611282;0.09569349961090001
\ssknn;0.4986556298227759;0.21938822370173644;0.435004995217599;0.08740299805101662
\knn;0.455520329378676;0.1715209857678258;0.3085932942477595;0.0930881316446834
\bpr;0.16177623131804436;0.103749344300167;0.6089993235179233;0.057553586574479995
\fpmc;0.07495055953902821;0.05082858168681307;F0.8115505990882192;0.020820599637440002
\fism;0.010516104887207247;0.00448385501369041;0.6241959329458042;F0.019743144613679996
\fossil;0.004799721366326474;0.0019460927441668086;0.6711711448375759;0.03429604765494
\end{filecontents*}
\DTLloaddb[noheader]{zalando}{tables_window_zalando-window-best.csv.txt}
		\begin{tabular}{@{}lrrrr@{}}
			\toprule
			\DTLforeach
			{zalando}
			{\key=Column1,\hit=Column2,\mrr=Column3,\cov=Column4,\pop=Column5}{%
				\ifthenelse{\value{DTLrowi}=1}{%
					\key & \mrr & \hit & \cov & \pop \\ \midrule
				}{%
					\key & \td{\mrr} & \td{\hit} & \td{\cov} & \td{\pop}
					\ifthenelse{\value{DTLrowi}=\DTLrowcount{rsc15}}{\\ \bottomrule }{\\ }%
				}%
			}%
		\end{tabular}
	\end{minipage}
	\newline
	\vspace{3pt}
	\newline
	\begin{minipage}[t][][b]{0.29\linewidth}
		\centering
		\scriptsize
		\vspace{17pt}
		\subcaption{RSC15-S}
		\label{tab:rsc15s_results}
		\DTLsetseparator{;}
		\begin{filecontents*}{tables_single_rsc15s-20.csv.txt}
Dataset;RSC15;A
Metric;HR@20;MRR@20
\gru;F0.719;F0.3119
\smf;0.7126;0.308673031
\sr;0.690328813;0.307548677
\vsknn;0.675211993;0.273589021
\sknn;0.641096283;0.250374604
\end{filecontents*}
\DTLloaddb[noheader]{rsc15s}{tables_single_rsc15s-20.csv.txt}
		\begin{tabular}{@{}lrrr@{}}
			\toprule
			\DTLforeach{rsc15s}
			{\key=Column1,\hr=Column2,\mrr=Column3}{%
				\ifthenelse{\value{DTLrowi}=1}{%
					\key & & \multicolumn{2}{c}{\hr}
				}{%
					\ifthenelse{\value{DTLrowi}=2}{%
						\key & & \mrr & \hr
					}{%
						\key & & \td{\mrr} & \td{\hr}
					}%
				}%
				\ifthenelse{\value{DTLrowi}=2}{%
					\\ \midrule
				}{%
					\ifthenelse{\value{DTLrowi}=\DTLrowcount{rsc15s}}{%
						\\ \bottomrule
					}{%
						\\
					}%
				}%
			}%
		\end{tabular}
	\end{minipage}
	\begin{minipage}[t]{0.69\linewidth}
		\centering
		\scriptsize
		\caption{Precision (P@20) and Recall (R@20) for the e-commerce datasets. The rows are ordered by the P@20 values for the \emph{TMALL} data set, which led to a relatively consistent ranking of the algorithms.}
		\label{tab:results_ecommerce_prm}
		\DTLsetseparator{;}
		\begin{filecontents*}{tables_precision-recall_ecommerce-20-new.csv.txt}
Dataset;RSC15;A;TMALL;A;ROCKET;A;ZALANDO;A
Metric;P@20;R@20;P@20;R@20;P@20;R@20;P@20;R@20
\sknn;0.0856;0.4638;F0.0945;F0.3122;F0.0562;F0.4779;0.0742;0.2018
\vsknn;F0.092;0.4935;0.0876;0.2907;0.0553;0.462;F0.0757;F0.207
\smf;0.0916;F0.5005;0.0679;0.2304;0.0471;0.3971;0.0616;0.1752
\gru;0.0853;0.4698;0.0677;0.2331;0.0458;0.3997;0.0646;0.1814
\sr;0.0889;0.4876;0.0519;0.1928;0.038;0.3417;0.0595;0.1744
\end{filecontents*}
\DTLloaddb[noheader]{precommerce}{tables_precision-recall_ecommerce-20-new.csv.txt}
		\begin{tabular}{@{}lrrrrrrrrrrrr@{}}
			\toprule
			\DTLforeach{precommerce}
			{\key=Column1,\pa=Column2,\ra=Column3,\pb=Column4,\rb=Column5,\pc=Column6,\rc=Column7,\pd=Column8,\rd=Column9}{%
				\ifthenelse{\value{DTLrowi}=1}{%
					\key & & \multicolumn{2}{c}{\pa} & & \multicolumn{2}{c}{\pb} & & \multicolumn{2}{c}{\pc} & &  \multicolumn{2}{c}{\pd}
				}{%
					\ifthenelse{\value{DTLrowi}=2}{%
						\key & & \pa & \ra & & \pb & \rb & & \pc & \rc & & \pd & \rd
					}{%
						\key & & \td{\pa} & \td{\ra} & & \td{\pb} & \td{\rb} & & \td{\pc} & \td{\rc} & & \td{\pd} & \td{\rd}
					}%
				}%
				\ifthenelse{\value{DTLrowi}=2}{%
					\\ \midrule
				}{%
					\ifthenelse{\value{DTLrowi}=\DTLrowcount{precommerce}}{%
						\\ \bottomrule
					}{%
						\\
					}%
				}%
			}%
		\end{tabular}
	\end{minipage}
\end{table}

\subsubsection{Accuracy Measures}
The results when the task is to predict the \emph{immediate next} element in a session (as done in \citep{Hidasi2017,JannachLudewig2017RecSys}) are shown in Tables \ref{tab:rsc15_results} to \ref{tab:rsc15s_results}. The following observations in terms of the hit rate and the MRR can be made.\footnote{We provide additional results that were obtained for measurements taken at multiple list lengths in Appendix \ref{appx:full_results}.}

\begin{itemize}
\item The lowest accuracy values are almost consistently achieved across all datasets by the family of Factorized Markov Chain approaches (\fism, \fpmc and \fossil) and the session-aware \bpr variant. \bpr in fact often exhibits the best performance among these methods even though it was not designed for sequential recommendation problems.
In two cases, however, the session-based \bpr variant led to very competitive results when the measurement was taken at a recommendation list length of 1, although at the potential price of a high popularity bias and low coverage.
Apart from this phenomenon, our results indicate that the methods that were designed under the assumption of longer-term and richer user profiles are often not particularly well suited for the specifics of session-based recommendation problems.
\item The simple pairwise association methods (\ar and \sr) mostly occupy the middle places in our comparison. In most cases, it is preferable to consider the available sequentiality information (\sr). Only for the \emph{TMALL} dataset, where the transactions of an entire day are considered as a session\footnote{In the dataset, timestamps are only available at the granularity of days.}, and for \emph{RETAILR}, the sequence-agnostic \ar method is slightly better in terms of the hit rate.
In terms of the overall ranking, the trivial \sr method is, to some surprise, among the \emph{top-performing} methods for two of the datasets in terms of the MRR, with good results also for the hit rate. The \mc method finally, is usually placed somewhere in the middle of the ranking. Similar to the \sr method, it is very strong in terms of the MRR for two of the datasets.
\item The performance of the newly-proposed \smf method is very strong for the \emph{RSC15} and \emph{RSC15-S} dataset and in the middle ranges for the other datasets. The \smf method consistently outperforms the factorization-based methods from the literature, apparently due to the embedding of the current user session.
\item \gru is consistently among the top five algorithms in this comparison in terms of the hit rate and exhibits competitive performance results also with respect to the MRR. The method is outperforming all other methods on the \emph{RSC15(-S)} datasets in terms of the hitrate and is competitive w.r.t.~the MRR, where the differences between the top-performing methods are tiny.
On the other datasets, the accuracy results of \gru are, however, often significantly lower than those of the best-performing methods.\footnote{We applied the Wilcoxon signed-rank test ($\alpha=0.05$) to determine the significance of differences between the two best performing approaches for each dataset.}
\item For each of the datasets, one of the proposed neighborhood-based methods was usually the winner in terms of the hit rate and the MRR (except for \emph{RSC15(-S)} and the MRR on \emph{ZALANDO}). Using one of the variants that considers sequentiality information is usually favorable, except for the case of the \emph{TMALL} dataset. The most consistent performance of the neighborhood-based methods is achieved with the \vsknn method which uses a specific sequence-aware similarity measure that gives more weight to the most recent interactions. Generally, the results suggest that there is even room for further improvement in the context of the neighborhood-based methods. In the experiments reported in this work, we could, for example, observe that using a slightly different similarity measure already led to substantial performance improvements for some of the datasets.
\end{itemize}

\paragraph{Precision and Recall for the Remaining Session.}
The ranking of the \emph{best-performing algorithms} when evaluating \emph{all} subsequent elements of a session (not only the immediate next click) and measuring precision and recall is given in Table \ref{tab:results_ecommerce_prm}. We report the detailed results for all algorithms for all measurements in the appendix.

The obtained results are mostly in line with the previously reported observations. The best performance is achieved by the neighborhood-based methods, with \vsknn working very well across all datasets. Differently from the previous measurement, \gru shows a lower performance for the \emph{RSC-15} dataset than the other methods. This is probably due to the fact that \gru is optimized to predict the \emph{immediate} next action.
Generally, which type of accuracy measurement---focusing on the prediction of the immediate next element or considering the prediction of any item that is relevant in the session as a success---is more appropriate, depends on the application domain. Our results show that the kNN-based methods are successful in both forms, i.e., they are often good at predicting the next element while, at the same time, they many times include \emph{more} items that are relevant for the given session than, e.g., \gru.

\begin{figure}[!t]
	\centering
	\begin{subfigure}[b]{0.5\textwidth}
		\centering
		\begin{tikzpicture}
		\small
		\begin{axis}[
		xlabel={ List length },
		x dir=reverse,
		legend style={
			at={(current bounding box.south-|current axis.south)},
			anchor=north,
			legend columns=2
		},
		legend entries = {\sknn, \vsknn, \gru, \sr},
		xmajorgrids=true,
		ymajorgrids=true,
		grid style=dashed,
		width=1.1\textwidth,
		height=0.6\textwidth,
		cycle list name = color list
		]
		\addplot[color1] table[x=List Length,y=S-KNN,col sep=semicolon] {charts_tmall-hitrate.csv.txt};
		\addplot[color2] table[x=List Length,y=SV-KNN,col sep=semicolon] {charts_tmall-hitrate.csv.txt};
		\addplot[color3] table[x=List Length,y=GRU,col sep=semicolon] {charts_tmall-hitrate.csv.txt};
		\addplot[color4] table[x=List Length,y=SR,col sep=semicolon] {charts_tmall-hitrate.csv.txt};
		\end{axis}
		\end{tikzpicture}
		\caption{TMALL}
		\label{fig:hitrate-tmall-hr}
	\end{subfigure}%
	\begin{subfigure}[b]{0.5\textwidth}
		\centering
		\begin{tikzpicture}
		\small
		\begin{axis}[
		xlabel={ List length },
		x dir=reverse,
		legend style={
			at={(current bounding box.south-|current axis.south)},
			anchor=north,
			legend columns=2
		},
		legend entries = {\sknn, \vsknn, \gru, \sr},
		xmajorgrids=true,
		ymajorgrids=true,
		grid style=dashed,
		width=1.1\textwidth,
		height=0.6\textwidth,
		cycle list name = color list
		]
		\addplot[color1] table[x=List Length,y=S-KNN,col sep=semicolon] {charts_retailrocket-hitrate.csv.txt};
		\addplot[color2] table[x=List Length,y=SV-KNN,col sep=semicolon] {charts_retailrocket-hitrate.csv.txt};
		\addplot[color3] table[x=List Length,y=GRU,col sep=semicolon] {charts_retailrocket-hitrate.csv.txt};
		\addplot[color4] table[x=List Length,y=SR,col sep=semicolon] {charts_retailrocket-hitrate.csv.txt};
		\end{axis}
		\end{tikzpicture}
		\caption{RETAILR}
		\label{fig:hitrate-retailrocket}
	\end{subfigure}
	\vspace{-10pt}
	\caption{Hit rate (HR) for two of the e-commerce datasets when reducing the recommendation list length from 20 to 1.}
\end{figure}
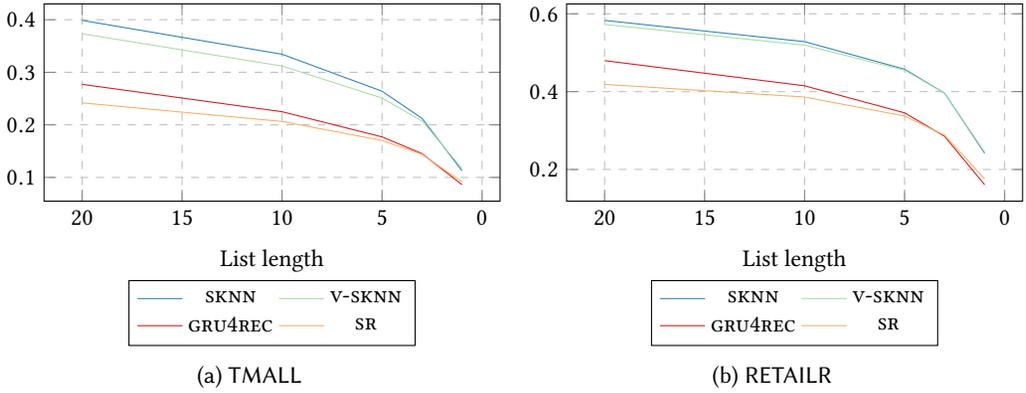

\paragraph{Impact of Different List Lengths}
To see if the recommendation list length at which the measurement is taken has an influence on the algorithm ranking, we varied the length from 20 to 1. Figure \ref{fig:hitrate-tmall-hr} and Figure \ref{fig:hitrate-retailrocket} show how the best algorithms perform for the \emph{TMALL} and \emph{RETAILR} datasets when different list lengths are used in the evaluation. The results show that the ranking of the algorithms can in fact be affected by the change of the list length.

Specifically, the differences between the nearest-neighbor methods and the \gru and \sr methods becomes gradually smaller for shorter list lengths. This is not too surprising because both \gru and \sr focus on the prediction of the immediate next action and often lead to better performance values in terms of the MRR. Since the particular evaluation protocol here also only focuses on the correct prediction of the next item, the effect might however be overemphasized due to the specific measurement method. An interesting observation is that at list length 1, \bpr and to some extent the \fpmc method lead to the best results for some e-commerce datasets. In the case of \bpr, this however comes at the price of a high popularity tendency of the algorithm and a comparably low coverage (see Table \ref{tab:appendix_window_tmall_results} in Appendix \ref{appx:full_results}).


\subsubsection{Cold-Start and Sparsity Effects}
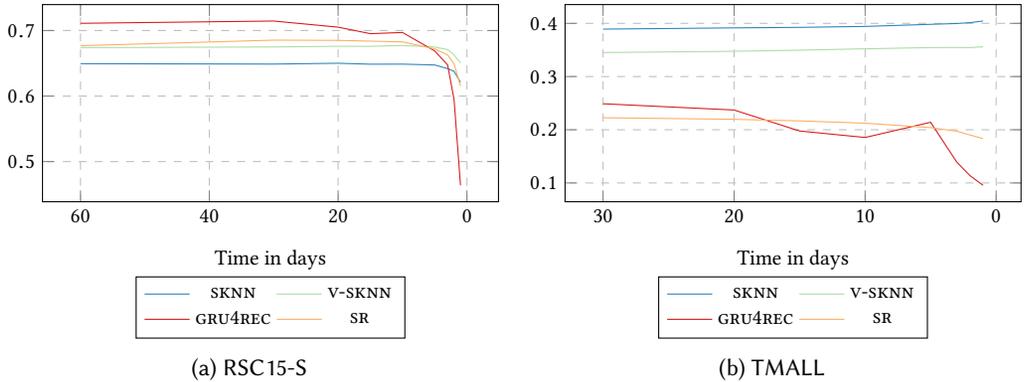
\begin{figure}[!ht]
	\centering
	\begin{subfigure}[b]{0.5\textwidth}
		\centering
		\begin{tikzpicture}
		\footnotesize
		\begin{axis}[
		xlabel={ Time in days },
		x dir=reverse,
		legend style={
			at={(current bounding box.south-|current axis.south)},
			anchor=north,
			legend columns=2
		},
		legend entries = {\sknn, \vsknn, \gru, \sr},
		xmajorgrids=true,
		ymajorgrids=true,
		grid style=dashed,
		width=1.1\textwidth,
		height=0.6\textwidth,
		cycle list name = color list
		]
		\addplot[color1] table[x=Days,y=SKNN,col sep=semicolon] {charts_rsc15-sparsity.csv.txt};
		\addplot[color2] table[x=Days,y=V-SKNN,col sep=semicolon] {charts_rsc15-sparsity.csv.txt};
		\addplot[color3] table[x=Days,y=GRU4REC,col sep=semicolon] {charts_rsc15-sparsity.csv.txt};
		\addplot[color4] table[x=Days,y=SR,col sep=semicolon] {charts_rsc15-sparsity.csv.txt};
		\end{axis}
		\end{tikzpicture}
		\caption{RSC15-S}
		\label{fig:trainsize-rsc15-hr}
	\end{subfigure}%
	\begin{subfigure}[b]{0.5\textwidth}
		\centering
		\begin{tikzpicture}
		\footnotesize
		\begin{axis}[
		xlabel={ Time in days },
		x dir=reverse,
		legend style={
			at={(current bounding box.south-|current axis.south)},
			anchor=north,
			legend columns=2
		},
		legend entries = {\sknn, \vsknn, \gru, \sr},
		xmajorgrids=true,
		ymajorgrids=true,
		grid style=dashed,
		width=1.1\textwidth,
		height=0.6\textwidth,
		cycle list name = color list
		]
		\addplot[color1] table[x=Days,y=SKNN,col sep=semicolon] {charts_tmall-sparsity.csv.txt};
		\addplot[color2] table[x=Days,y=V-SKNN,col sep=semicolon] {charts_tmall-sparsity.csv.txt};
		\addplot[color3] table[x=Days,y=GRU4REC,col sep=semicolon] {charts_tmall-sparsity.csv.txt};
		\addplot[color4] table[x=Days,y=SR,col sep=semicolon] {charts_tmall-sparsity.csv.txt};
		\end{axis}
		\end{tikzpicture}
		\caption{TMALL}
		\label{fig:trainsize-tmall-hr}
	\end{subfigure}
	\vspace{-10pt}
	\caption{HR@20 for two e-commerce datasets when artificially reducing the size of the training set from 60 days to 1 day.}
\end{figure}

Previous experiments on the \emph{RSC15} dataset revealed that discarding major parts of the older data has no strong impact on the prediction accuracy, at least in the e-commerce domain \citep{JannachLudewig2017RecSys}. We therefore made additional experiments to analyze the effects in more depth.
Figure \ref{fig:trainsize-rsc15-hr} and Figure \ref{fig:trainsize-tmall-hr} show the results of this simulation for two of the e-commerce datasets.

The results for the \emph{RSC15-S} (single-split) dataset (Figure \ref{fig:trainsize-rsc15-hr}) are in line with what was previously reported in \citep{JannachLudewig2017RecSys}. In the e-commerce domain, the user behavior seems to be strongly influenced by recent sales trends, an effect that was also reported in \citep{JannachLudewigLerche2017umuai}. Discarding most of the historical data has almost no influence on the resulting hit rates. This behavior is similar for all compared algorithms. Only in the extreme case when only the data of the last few days is considered, the performance of the algorithms degrades. A similar observation can be made for the \emph{TMALL} dataset. Generally, the observations also explain why the recency-based neighborhood sampling approach implemented in the kNN methods does not have a strong negative effect on the accuracy. In fact, focusing on the most recent sessions when looking for similar neighbors has shown to have a positive effect in \citep{JannachLudewig2017RecSys}, when compared to a random neighborhood selection scheme.

\paragraph{Considering other Types of Events}
In the reported experiments, the models were trained with past item view events and we also predicted the next view event for a given session beginning. This choice was made to make our work comparable with previous research. Additional types of events (e.g., ``add-to-wishlist'', ``add-to-cart'') can easily incorporated as positive preference signals into the investigated algorithms. How to weight the different types of signals nd how to interpret signals like ``remove-from-wishlist'' is an area for future research.

Depending on the domain, also different types of events might be in the focus as well in the prediction phase. In the experiments reported here, we predict item views. In our previous research on the topic \citep{JannachLudewig2017RecSys}, we also made experiments in which we focused on the prediction of purchase events. In these experiments, the ranking of the algorithms was similar for the item prediction and the purchase prediction tasks. However, some previous research suggests that view-based collaborative filtering algorithms lead to sometimes quite different recommendations than purchase-based ones and also differ in their effectiveness \citep{DBLP:conf/icis/LeeH14}. In general, the choice of the prediction target should therefore be made with the goal of the recommender in mind, e.g., increase user attention and click-through-rates vs.~increasing sales, see, e.g.,  \citep{JannachHegelich2009}.

\subsubsection{Coverage and Popularity Bias}
The results listed in Table \ref{tab:rsc15_results} to Table \ref{tab:zalando_results} show that in terms of the coverage (or: aggregate diversity), the factorization-based methods consistently lead to the highest values, i.e., they place the largest number of different items into the top-n lists of the users.
\gru represents in all datasets, except \emph{RETAILR}, the other extreme and seems to focus its recommendations on a comparably narrow range of items. In particular in the case of the \emph{TMALL} dataset, the coverage of the item space of \gru is as low as 0.15, i.e., the top-20 recommendations for all given sessions in the test set cover only 15\,\% of the available items. To what extent low coverage is undesired, again depends on the specific application domain.

Not many consistent patterns can be identified with regard to the popularity biases of the different algorithms. \bpr, as was previously discussed by \cite{JannachLercheEtAl2015}, has a comparably strong tendency to focus on generally popular items. Our newly proposed \smf method exhibits a similar tendency across all datasets.
The \fpmc method usually represents the other end of the spectrum. The tendency of the many of the other algorithms to recommend popular items seems to strongly depend on the dataset characteristics. According to our previous work \citep{JannachLudewig2017RecSys}, the basic \sknn method tends to recommend slightly more popular items than \gru. In this new series of measurements, this is, however, not consistently the case across the datasets.

\subsection{Media Datasets}
Table \ref{tab:music-results},  Table \ref{tab:results_music_prm}, and Table \ref{tab:clef_results} show the results for the music and news domains, respectively.

\begin{table}
	\setlength\tabcolsep{1.8pt}
	\caption{Hit rate (HR), Mean reciprocal rank (MRR), catalog coverage (COV), and the average popularity (POP) for a list length of 20 tested on the music datasets. The tables show the top ten algorithms ordered by MRR@20. The best results are highlighted and significant differences are marked with a star.}
	\label{tab:music-results}
	\begin{minipage}[t]{0.5\linewidth}
		\scriptsize
		\centering
		\subcaption{NOWPLAYING}
		\label{tab:nowplaying_results}
		\DTLsetseparator{;}
		\begin{filecontents*}{tables_window_nowplaying-window-best-small.csv.txt}
Metrics;HR@20;MRR@20;COV@20;POP@20
\sr;0.20330093004137612;F0.10528233338116774;0.4655552558621235;F0.024730176598142573
\gru;0.19697080994515517;0.10183773881491417;0.4331073098032056;0.051595999321560004
\mc;0.15819295309199927;0.09710373027839683;0.2935597662910592;0.028392314499010345
\sfsknn;0.16472102242912584;0.09542545958337992;0.2772706209662744;0.031124818344788863
\smf;0.18250350850623123;0.0881973921533314;0.24169590926477466;0.09158628256924001
\vsknn;0.25519534027400326;0.07846291182049828;0.4282988814609845;0.06394308191608
\ssknn;S0.2621751626538865;0.07767932948878144;0.4148821745113597;0.06239017890151385
\ar;0.20760597984068846;0.07101582651548904;0.4530542658544617;0.05112347788421203
\knn;0.24293773240447333;0.06892723891887671;0.3006939930271181;0.06904753925328479
\iknn;0.1821707958623518;0.05691473037572496;F0.5798533813861946;0.029386653672800002
\end{filecontents*}
\DTLloaddb[noheader]{nowp}{tables_window_nowplaying-window-best-small.csv.txt}
		\begin{tabular}{@{}lrrrr@{}}
			\toprule
			\DTLforeach
			{nowp}
			{\key=Column1,\hit=Column2,\mrr=Column3,\cov=Column4,\pop=Column5}{%
				\ifthenelse{\value{DTLrowi}=1}{%
					\key & \mrr & \hit & \cov & \pop \\ \midrule
				}{%
					\key & \td{\mrr} & \td{\hit} & \td{\cov} & \td{\pop}
					\ifthenelse{\value{DTLrowi}=\DTLrowcount{nowp}}{\\ \bottomrule }{\\ }%
				}%
			}%
		\end{tabular}
	\end{minipage}
	\begin{minipage}[t]{0.49\linewidth}
		\sisetup{
			round-mode          	= places, 
			round-precision     	= 4, 
			detect-weight			= true,
			detect-inline-weight	= math
		}
		\scriptsize	
		\centering
		\subcaption{8TRACKS}
		\label{tab:8tracks_results}
		\DTLsetseparator{;}
		\begin{filecontents*}{tables_window_8tracks-window-best-small.csv.txt}
Metrics;HR@20;MRR@20;COV@20;POP@20
\ar;0.025537226641832632;S0.0071344678950149305;0.4529592805558593;0.09124478778517413
\smf;0.023097358668468928;0.006410840258113717;0.15279806012504943;0.08644607581844
\sr;0.017078491043338327;0.006388254375150318;0.49671191415575483;0.053105777351315574
\sfsknn;0.011875803132067953;0.0063729917038662836;0.3049360436400458;0.03628372678490173
\vsknn;0.035221310860267294;0.005728747787049435;0.40805756552397277;0.1194310461902
\knn;S0.037553230817024434;0.005346597961362094;0.24308137719690012;0.107966569803515
\iknn;0.017672410010551927;0.005057727627289346;F0.6956070997367665;F0.02453855201092
\gru;0.018948046788232663;0.005006143973289785;0.0692959309197823;0.12229408310439997
\ssknn;0.029308635707672948;0.004759648408507964;0.4509183917877343;0.08068713619618947
\mc;0.009864966139086032;0.004647327876972722;0.34961802750366466;0.032067180087403245
\end{filecontents*}
\DTLloaddb[noheader]{8tracks}{tables_window_8tracks-window-best-small.csv.txt}
		\begin{tabular}{@{}lrrrr@{}}
			\toprule
			\DTLforeach
			{8tracks}
			{\key=Column1,\hit=Column2,\mrr=Column3,\cov=Column4,\pop=Column5}{%
				\ifthenelse{\value{DTLrowi}=1}{%
					\key & \mrr & \hit & \cov & \pop \\ \midrule
				}{%
					\key & \td{\mrr} & \td{\hit} & \td{\cov} & \td{\pop}
					\ifthenelse{\value{DTLrowi}=\DTLrowcount{8tracks}}{\\ \bottomrule }{\\ }%
				}%
			}%
		\end{tabular}
	\end{minipage}
	\newline
	\begin{minipage}[t]{0.5\linewidth}
		\sisetup{
			round-mode          	= places, 
			round-precision     	= 3, 
			detect-weight			= true,
			detect-inline-weight	= math
		}
		\scriptsize
		\centering
		\subcaption{30MUSIC}
		\label{tab:30music_results}
		\DTLsetseparator{;}
		\begin{filecontents*}{tables_window_30music-window-best-small.csv.txt}
Metrics;HR@20;MRR@20;COV@20;POP@20
\sr;0.33231044108669805;S0.23772052003287025;0.38925517906143803;0.023191889952974304
\mc;0.2843762506563592;0.23181348091362727;0.2038404905078981;F0.02052176494803105
\gru;0.3256921930475206;0.22637070678996438;0.34473947064448723;0.055626852402360004
\sfsknn;0.2856080354156377;0.20793384008961704;0.18542857832931245;0.021948040357426674
\smf;0.2842852312902921;0.1776513453779502;0.15084817778241272;0.10477503097518
\vsknn;0.38186710488868536;0.10987333182723558;0.3169604766160029;0.053790465016919996
\iknn;0.29710759799021424;0.10861031605631219;F0.4595781931348313;0.02259070839078
\ssknn;S0.38558615148401515;0.10766464446940316;0.2931429222870193;0.051544103102565686
\ar;0.3087878615598692;0.09600482773206484;0.3524256883071061;0.03935378819700435
\knn;0.34432121835286517;0.08979383429653967;0.19123264329097067;0.057422233743154395
\end{filecontents*}
\DTLloaddb[noheader]{30m}{tables_window_30music-window-best-small.csv.txt}
		\begin{tabular}{@{}lrrrr@{}}
			\toprule
			\DTLforeach
			{30m}
			{\key=Column1,\hit=Column2,\mrr=Column3,\cov=Column4,\pop=Column5}{%
				\ifthenelse{\value{DTLrowi}=1}{%
					\key & \mrr & \hit & \cov & \pop \\ \midrule
				}{%
					\key & \td{\mrr} & \td{\hit} & \td{\cov} & \td{\pop}
					\ifthenelse{\value{DTLrowi}=\DTLrowcount{30m}}{\\ \bottomrule }{\\ }%
				}%
			}%
		\end{tabular}
	\end{minipage}
	\begin{minipage}[t]{0.49\linewidth}
		\sisetup{
			round-mode          	= places, 
			round-precision     	= 4, 
			detect-weight			= true,
			detect-inline-weight	= math
		}
		\scriptsize
		\centering
		\subcaption{AOTM}
		\label{tab:aotm_results}
		\DTLsetseparator{;}
		\begin{filecontents*}{tables_window_aotm-window-best-small.csv.txt}
Metrics;HR@20;MRR@20;COV@20;POP@20
\smf;0.02977952259434848;F0.011114042873473774;0.24565864602206444;0.1998189436806
\sfsknn;0.014479746020294785;0.011067211602580989;0.3558828907082705;0.05083643631912734
\sr;0.019548723009244835;0.007661815074805208;0.5864089111558594;0.053318681123421155
\gru;0.015660913370266302;0.0071851526853896645;0.4652738424178652;0.11512799197800001
\mc;0.013283815491272291;0.006320633358865087;0.38032442126764143;0.04976425800839485
\ar;0.023328512880469783;0.005918018519036876;0.5531780611314797;0.10492135259939091
\vsknn;0.03778058768341828;0.0054883379160220154;0.536266152572802;0.1397131021964
\ssknn;0.03971644698481465;0.005472913138839716;0.5356801234472488;0.12892703313641266
\knn;S0.042948192003364784;0.005390000401035161;0.28021078406509725;0.16777852471003732
\iknn;0.018670846259286382;0.00490796614489057;F0.7879859817321977;F0.047286740480260006
\end{filecontents*}
\DTLloaddb[noheader]{aotm}{tables_window_aotm-window-best-small.csv.txt}
		\begin{tabular}{@{}lrrrr@{}}
			\toprule
			\DTLforeach
			{aotm}
			{\key=Column1,\hit=Column2,\mrr=Column3,\cov=Column4,\pop=Column5}{%
				\ifthenelse{\value{DTLrowi}=1}{%
					\key & \mrr & \hit & \cov & \pop \\ \midrule
				}{%
					\key & \td{\mrr} & \td{\hit} & \td{\cov} & \td{\pop}
					\ifthenelse{\value{DTLrowi}=\DTLrowcount{aotm}}{\\ \bottomrule }{\\ }%
				}%
			}%
		\end{tabular}
	\end{minipage}
	\newline
	\begin{minipage}[t]{1\linewidth}
		\sisetup{
			round-mode          	= places, 
			round-precision     	= 4, 
			detect-weight			= true,
			detect-inline-weight	= math
		}
		\centering
		\scriptsize
		\caption{Precision (P@20) and Recall (R@20) for the music datasets. The results are ordered by P@20 for \emph{8TRACKS}, which represents the largest music dataset in our evaluation.}
		\label{tab:results_music_prm}
		\DTLsetseparator{;}
		\begin{filecontents*}{tables_precision-recall_music-20.csv.txt}
Dataset;LFM;;8TRACKS;;30MUSIC;;AOTM;
Metric;P@20;R@20;P@20;R@20;P@20;R@20;P@20;R@20
\vsknn;F0.0717;F0.1909;F0.0122;0.0308;F0.1094;F0.2321;0.0133;0.0361
\sknn;0.068;0.1824;0.0117;F0.0313;0.1035;0.214;F0.0155;F0.044
\smf;0.0499;0.1453;0.0086;0.0218;0.0746;0.1655;0.0084;0.0259
\sr;0.0501;0.1465;0.0055;0.014;0.0878;0.201;0.0053;0.0146
\gru;0.0272;0.081;0.0037;0.0095;0.0404;0.0988;0.001;0.0027
\end{filecontents*}
\DTLloaddb[noheader]{prmusic}{tables_precision-recall_music-20.csv.txt}
		\begin{tabular}{@{}lrrrrrrrrrrrr@{}}
			\toprule
			\DTLforeach{prmusic}
			{\key=Column1,\pa=Column2,\ra=Column3,\pb=Column4,\rb=Column5,\pc=Column6,\rc=Column7,\pd=Column8,\rd=Column9}{%
				\ifthenelse{\value{DTLrowi}=1}{%
					\key & & \multicolumn{2}{c}{\pa} & & \multicolumn{2}{c}{\pb} & & \multicolumn{2}{c}{\pc} & & \multicolumn{2}{c}{\pd}
				}{%
					\ifthenelse{\value{DTLrowi}=2}{%
						\key & & \pa & \ra & & \pb & \rb & & \pc & \rc & & \pd & \rd
					}{%
						\key & & \td{\pa} & \td{\ra} & & \td{\pb} & \td{\rb} & & \td{\pc} & \td{\rc} & & \td{\pd} & \td{\rd}
					}%
				}%
				\ifthenelse{\value{DTLrowi}=2}{%
					\\ \midrule
				}{%
					\ifthenelse{\value{DTLrowi}=\DTLrowcount{precommerce}}{%
						\\ \bottomrule
					}{%
						\\
					}%
				}%
			}%
		\end{tabular}
	\end{minipage}
\end{table}
\begin{table}[h!t]
	\scriptsize
	\centering
	\caption{Hit rate (HR), Mean reciprocal rank (MRR), Precision (P), Recall (R), item coverage (COV), and average popularity (POP) results for a list length of 20 on the \emph{CLEF} dataset (ordered by MRR@20).}
	\label{tab:clef_results}
	\DTLsetseparator{;}
	\begin{filecontents*}{tables_clef-all-small.csv.txt}
Metrics;HR@20;MRR@20;COV@20;POP@20;P@20;R@20
\smf;0.706117067;F0.234423738;0.649565163;0.082843407;0.061677142;0.526821096
\vsknn;0.775557828;0.223575333;0.621024029;F0.082734924;0.067958935;0.58413196
\sr;0.671878104;0.222966353;F0.655356122;0.093288231;0.058275129;0.501955061
\gru;0.568499349;0.219950467;0.17442489;0.094286256;F0.072447866;F0.626009412
\knn;F0.778077591;0.21856583;0.612573028;0.084236067;0.065616015;0.577132601
\end{filecontents*}
\DTLloaddb[noheader]{clef}{tables_clef-all-small.csv.txt}
	\begin{tabular}{@{}lrrrrrr@{}}
		\toprule
		\DTLforeach{clef}
		{\key=Column1,\hit=Column2,\mrr=Column3,\p=Column4,\r=Column5,\cov=Column6,\pop=Column7}{%
			\ifthenelse{\value{DTLrowi}=1}{%
				\key & \mrr & \hit & \p & \r & \cov & \pop
			}{%
				\key & \td{\mrr} & \td{\hit} & \td{\p} & \td{\r} & \td{\cov} & \td{\pop}
			}%
			\ifthenelse{\value{DTLrowi}=1}{%
				\\ \midrule
			}{%
				\ifthenelse{\value{DTLrowi}=\DTLrowcount{clef}}{%
					\\ \bottomrule
				}{%
					\\
				}%
			}%
		}%
	\end{tabular}
\end{table}

\paragraph{Accuracy}
The accuracy results generally exhibit similar patterns as the results obtained for the e-commerce datasets. For these datasets, however, the winning strategy more strongly depends on the chosen measure. When the MRR is used as a performance measure, often the trivial baselines \sr or \ar lead to the best results. In terms of the hitrate, in contrast, usually one of the nearest neighbor methods again performs best.

With respect to the MRR measure also \gru exhibited very competitive performance, except for the \emph{8TRACKS} and \emph{AOTM} datasets, where the highest MRR values were achieved with the \ar and the \sknn method.
Looking at the playlist datasets (\emph{8TRACKS} and \emph{AOTM}), the comparably good results of the sequence-agnostic \ar and \sknn strategy indicate that the ordering of the tracks is not too important for the playlist creators. Among the neighborhood-based methods, \vsknn was again consistently among the top-performing methods. When looking at the standard precision and recall measurements for the five best-performing approaches in Table \ref{tab:results_music_prm}, we can see that \vsknn is the winning strategy across all datasets and that \gru is again less effective for this particular measurement.

Finally, looking at the \emph{news} domain, the average results shown in Table \ref{tab:clef_results} in general confirm the trends observed for the other datasets. The \vsknn method is top-performing on almost all measures. \gru also works comparably well on this dataset, especially on the precision and recall measures. Again, however, we can also observe a comparably low level of coverage and a comparably high tendency to recommend popular items.
In contrast to all other domains and datasets however, when looking at the results of the individual splits for the \emph{CLEF} dataset, we could observe that those are subject to large fluctuations. Depending on the day that was chosen for testing, the ranking of the algorithms in terms of the accuracy measures changes drastically, which we could not observe for any other dataset.

As mentioned in Section \ref{sec:evaluation-setup}, we conducted additional single split experiments to ensure that the reduced amount of training data in the sliding window protocol does not affect the performance of the model-based approaches.
The single-split results in Appendix \ref{appx:singleres} reveal \gru as the best-performing approach for this particular dataset, which was also the case for two of the individual splits. Thus, even though such large fluctuations did only occur in the news domain, this is an indicator that applying a single-split evaluation protocol can easily lead to ``random'' and misleading results.

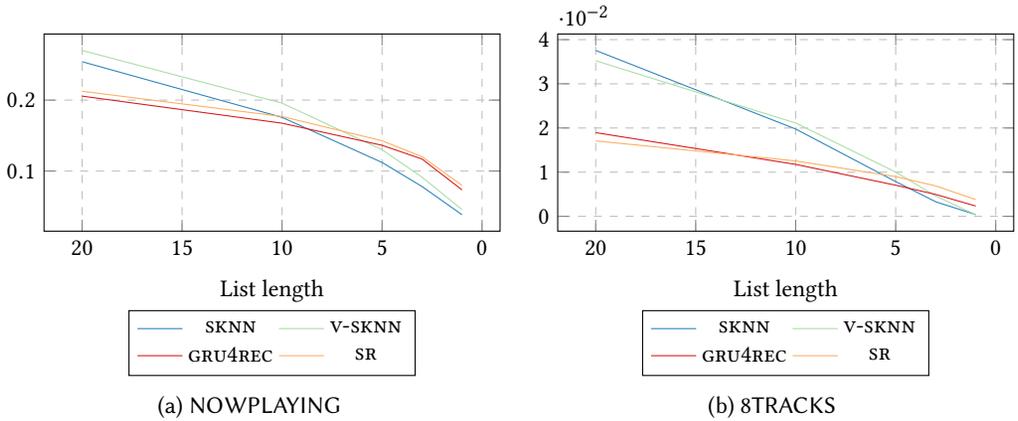
\begin{figure}[!t]
	\centering
	\begin{subfigure}[b]{0.5\textwidth}
		\centering
		\begin{tikzpicture}
		\small
		\begin{axis}[
		xlabel={ List length},
		x dir=reverse,
		legend style={
			at={(current bounding box.south-|current axis.south)},
			anchor=north,
			legend columns=2
		},
		legend entries = {\sknn, \vsknn, \gru, \sr},
		xmajorgrids=true,
		ymajorgrids=true,
		grid style=dashed,
		width=1.1\textwidth,
		height=0.6\textwidth,
		cycle list name = color list
		]
		\addplot[color1] table[x=List Length,y=S-KNN,col sep=semicolon] {charts_nowplaying-hitrate.csv.txt};
		\addplot[color2] table[x=List Length,y=SV-KNN,col sep=semicolon] {charts_nowplaying-hitrate.csv.txt};
		\addplot[color3] table[x=List Length,y=GRU,col sep=semicolon] {charts_nowplaying-hitrate.csv.txt};
		\addplot[color4] table[x=List Length,y=SR,col sep=semicolon] {charts_nowplaying-hitrate.csv.txt};
		\end{axis}
		\end{tikzpicture}
		\caption{NOWPLAYING}
		\label{fig:hitrate-nowplaying-hr}
	\end{subfigure}%
	\begin{subfigure}[b]{0.5\textwidth}
		\centering
		\begin{tikzpicture}
		\small
		\begin{axis}[
		xlabel={ List length },
		x dir=reverse,
		legend style={
			at={(current bounding box.south-|current axis.south)},
			anchor=north,
			legend columns=2
		},
		legend entries = {\sknn, \vsknn, \gru, \sr},
		xmajorgrids=true,
		ymajorgrids=true,
		grid style=dashed,
		width=1.1\textwidth,
		height=0.6\textwidth,
		cycle list name = color list
		]
		\addplot[color1] table[x=List Length,y=S-KNN,col sep=semicolon] {charts_8tracks-hitrate.csv.txt};
		\addplot[color2] table[x=List Length,y=SV-KNN,col sep=semicolon] {charts_8tracks-hitrate.csv.txt};
		\addplot[color3] table[x=List Length,y=GRU,col sep=semicolon] {charts_8tracks-hitrate.csv.txt};
		\addplot[color4] table[x=List Length,y=SR,col sep=semicolon] {charts_8tracks-hitrate.csv.txt};
		\end{axis}
		\end{tikzpicture}
		\caption{8TRACKS}
		\label{fig:hitrate-8tracks-hr}
	\end{subfigure}
	\vspace{-10pt}
	\caption{Hit rate (HR) for two of the music datasets when reducing the result list length from 20 to 1.}
\end{figure}

The effects when considering different list lengths for two of the datasets is shown in Figure \ref{fig:hitrate-nowplaying-hr} (\emph{NOWPLAYING}) and Figure \ref{fig:hitrate-8tracks-hr} (\emph{8TRACKS}). In contrast to the e-commerce datasets, the relative ranking of the algorithms even changes when the list lengths become shorter. For both datasets, the \gru method and the very simple \ar and \sr methods, respectively, are better in terms of the hit rate when it comes to very short list lengths. Considering the good results for the MRR for these methods (Table \ref{tab:nowplaying_results} and Table \ref{tab:8tracks_results}), this was expected. Again, the good performance of certain methods can be explained by the fact that these methods are optimized to predict the immediate next item of a given session.

\begin{figure}[!t]
	\centering
	\begin{subfigure}[b]{0.5\textwidth}
		\centering
		\begin{tikzpicture}
		\small
		\begin{axis}[
		xlabel={ Time in days },
		x dir=reverse,
		legend style={
			at={(current bounding box.south-|current axis.south)},
			anchor=north,
			legend columns=2
		},
		legend entries = {\sknn, \vsknn, \gru, \sr},
		xmajorgrids=true,
		ymajorgrids=true,
		grid style=dashed,
		width=1.1\textwidth,
		height=0.6\textwidth,
		cycle list name = color list
		]
		\addplot[color1] table[x=Days,y=SKNN,col sep=semicolon] {charts_8tracks-sparsity-new.csv.txt};
		\addplot[color2] table[x=Days,y=V-SKNN,col sep=semicolon] {charts_8tracks-sparsity-new.csv.txt};
		\addplot[color3] table[x=Days,y=GRU4REC,col sep=semicolon] {charts_8tracks-sparsity-new.csv.txt};
		\addplot[color4] table[x=Days,y=SR,col sep=semicolon] {charts_8tracks-sparsity-new.csv.txt};
		\end{axis}
		\end{tikzpicture}
		\caption{8TRACKS}
		\label{fig:trainsize-8t-hr}
	\end{subfigure}%
	\begin{subfigure}[b]{0.5\textwidth}
		\centering
		\begin{tikzpicture}
		\small
		\begin{axis}[
		xlabel={ Time in days },
		x dir=reverse,
		legend style={
			at={(current bounding box.south-|current axis.south)},
			anchor=north,
			legend columns=2
		},
		legend entries = {\sknn, \vsknn, \gru, \sr},
		xmajorgrids=true,
		ymajorgrids=true,
		grid style=dashed,
		width=1.1\textwidth,
		height=0.6\textwidth,
		cycle list name = color list
		]
		\addplot[color1,restrict x to domain =3:100] table[x=Days,y=SKNN,col sep=semicolon] {charts_nowplaying-sparsity.csv.txt};
		\addplot[color2,restrict x to domain =3:100] table[x=Days,y=V-SKNN,col sep=semicolon] {charts_nowplaying-sparsity.csv.txt};
		\addplot[color3,restrict x to domain =3:100] table[x=Days,y=GRU4REC,col sep=semicolon] {charts_nowplaying-sparsity.csv.txt};
		\addplot[color4,restrict x to domain =3:100] table[x=Days,y=SR,col sep=semicolon] {charts_nowplaying-sparsity.csv.txt};
		\end{axis}
		\end{tikzpicture}
		\caption{NOWPLAYING}
		\label{fig:trainsize-nowplaying-hr}
	\end{subfigure}
	\vspace{-10pt}
	\caption{HR@20 for two music datasets when incrementally reducing the size of the training set to 1 day.}
\end{figure}
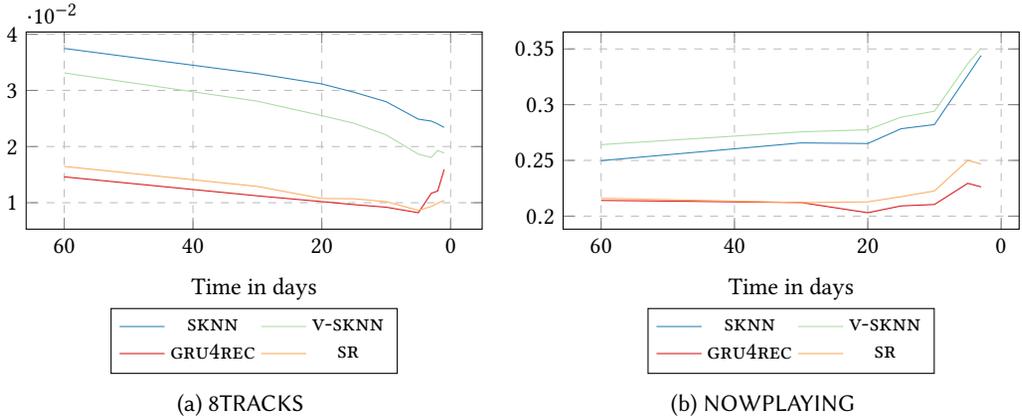

\paragraph{Cold-Start and Sparsity Effects}
An interesting effect can be observed when older data is discarded to simulate sparsity effects. Figure \ref{fig:trainsize-8t-hr} and Figure \ref{fig:trainsize-nowplaying-hr} show the results for the \emph{8TRACKS} and \emph{NOWPLAYING} datasets, respectively.\footnote{The other media datasets did not exhibit any notable particularities.} While for the \emph{8TRACKS} playlist dataset the accuracy values more or less consistently decrease when older data is discarded, we can observe an \emph{increase} in accuracy for the \emph{NOWPLAYING} dataset. Remember that this dataset is based on the analysis of user posts on Twitter about their current listening behavior. Obtaining the highest accuracy values when only considering the very last days means that this dataset is strongly dominated by short-term popularity trends and that the recommendation of older, non-trending tracks is detrimental to the accuracy results.

\paragraph{Coverage and Popularity Bias}
In terms of coverage (see Table \ref{tab:music-results}), the findings for
datasets from the media domain are also mostly in line with those for the e-commerce datasets. The ranking of the algorithms varies largely across the datasets. The differences are, however, often less pronounced. Regarding the popularity tendency of the algorithms, methods that are based on pairwise sequences (\sr and \mc) in most cases lead to the recommendation of lesser known items, while nearest-neighbor-based techniques quite often focus on the recommendation of comparably popular objects.

\subsection{Computational Complexity \& Memory Usage}
The methods included in our comparison vary largely in terms of the computational complexity and their memory requirements. Since neighborhood-based methods do not scale well when applied in a naive manner, we used implementation variants that rely on neighborhood sampling and specific in-memory data structures. The comparison of \sknn method and \gru in \citep{JannachLudewig2017RecSys} showed that, with such an implementation, recommendations can be quickly computed at prediction time with nearest neighbor methods, even though the prediction performance of model-based techniques like \gru could not be achieved.

To enable comparability with previous research \citep{JannachLudewig2017RecSys}, we report the running times and memory demands for the single-split \emph{RSC15-S} dataset, which is also the largest one in terms of the recorded user actions. Additionally, we include the \emph{8TRACKS} dataset, which is rather small compared to \emph{RSC15-S} in terms of the number of events, but has the largest product catalog of all datasets. Table \ref{tab:computational-complexity} shows the times required for training the model (if applicable), the time needed to compute a recommendation at prediction time, and the memory requirements for the internal data structures. The reported results were obtained when using an Intel Core i7 4790K processor with 32GB of DDR3-1600 memory and a Nvidia GeForce GTX 960 graphics card with 2GB of memory. The following observations can be made.

\sisetup{
	round-mode          = places, 
	round-precision     = 2, 
}
\begin{table}[!t]
	\scriptsize
	\centering
	\caption{Overview of Computation Times and Memory Requirements for the \emph{RSC15-S} dataset and the first split of the \emph{8TRACKS} dataset, ordered in terms of required training times for the \emph{RSC15-S} dataset.}
	\label{tab:computational-complexity}
	\DTLsetseparator{;}
	\begin{filecontents*}{tables_speed-size_speed-size.csv.txt}
Dataset;RSC15-S;;;8TRACKS;;
Technique/Metric;Train. (min);Pred. (ms);Mem. (MB);Train.;Pred.;Mem.
\mc;0.766562247;3.343721757;38.45091248;0.053311714;14.70514048;143.5811081
\knn;1.237669567;33.05183324;6051.389793;0.041577093;52.96092221;352.8337326
\ssknn;1.264935525;30.25657467;6051.389908;0.048223988;51.00724298;352.833847
\vsknn;1.296913032;32.67073917;6051.390427;0.043674552;52.43400632;352.8343658
\sfsknn;1.724094367;29.81706039;6253.91909;0.063741318;18.81878686;493.1173019
\sr;2.4143616;3.136563265;54.3161087;0.175828815;15.88340133;283.6713409
\ar;3.004836698;3.357926236;40.06252289;0.281904769;15.78716634;257.2489395
\fism;356.8410775;8.404075614;4936.682861;35.07045571;60.72097916;386.7705536
\gru (GPU);385.3518352;7.432489252;59.17201233;19.07750254;58.0759628;587.578186
\bpr;392.6095387;8.365201914;8009.314171;42.69282834;65.14561575;771.4822769
\smf (GPU);446.6671667;14.01939527;1639.984871;77.49882066;43.36411324;1805.347778
\fpmc;469.3924318;9.075076481;6786.366295;60.92341224;71.58162675;1301.512657
\fossil;499.1907578;10.55595631;4986.518127;50.99271194;64.9063558;581.5421982
\end{filecontents*}
\DTLloaddb[noheader]{speedsize}{tables_speed-size_speed-size.csv.txt}
	\begin{tabular}{@{}lrrrrrrr@{}}
		\toprule
		\DTLforeach*{speedsize}
		{\key=Column1,\ta=Column2,\pa=Column3,\ma=Column4,\tb=Column5,\pb=Column6,\mb=Column7}{%
			\ifthenelse{\value{DTLrowi}=1 \or \value{DTLrowi}=2}{%
				\ifthenelse{\value{DTLrowi}=1}{%
					\key & \multicolumn{3}{c}{\ta} & & \multicolumn{3}{c}{\tb}\\
				}{%
					\key & \ta & \pa & \ma & & \tb & \pb & \mb\\ \midrule
				}%
			}{%
				\key & \tdn{\ta} & \tdn{\pa} & \tdz{\ma} & & \tdn{\tb} & \tdn{\pb} & \tdz{\mb}
				\ifthenelse{\value{DTLrowi}=\DTLrowcount{speedsize}}{%
					\\ \bottomrule
				}{%
					\\
				}%
			}%
		}%
	\end{tabular}
\end{table}

\paragraph{Running Times}
The simple methods in our comparative evaluation need from less than one to about three minutes of ``training'' (e.g., co-occurrence counting or in-memory data structure setup) for the \emph{RSC15-S} dataset. The factorization-based methods and the deep learning based method, on the other hand, need about 6 to 8 hours to learn a model for the single data split.
Note that while the deep learning method \gru and the factorization-based approach \smf do not take the longest absolute time in this comparison, they are the only method for which the computations are done on the GPU. Running \gru, for example, on a CPU tripled the computation times according to the measurements in \citep{JannachLudewig2017RecSys}.

Looking at the times needed to compute a single recommendation list, given a session beginning, we can observe that the simple rule-based methods \ar, \mc, and \sr are among the fastest ones with prediction times at about 3 ms for the \emph{RSC15-S} dataset. The factorization-based methods and \gru are also very efficient, with prediction times mostly below 10 ms on average. The nearest-neighbor methods are slower for this task as they have to consider the neighbors in the prediction process. Since the neighbors can be determined through fast lookup operations, the overall prediction time even for the more elaborate \ssknn and \vsknn similarity schemes never exceeds 33 ms for creating a recommendation list.

Looking at the \emph{8TRACKS} dataset with its large number of items, we can, however, see that the prediction times for many algorithms, including \gru and several of the factorization-based ones significantly increase, while the prediction time for the neighborhood models only doubles. In the end, making the neighborhood-based computations is at least as fast as computing the predictions based on the offline-trained models. Overall, due to the used in-memory data structures and through the neighborhood sampling approach, such neighborhood models are also suited under the narrow time constraints of real-time recommendations. Differently from other methods, newly arriving interaction signals can be easily included in the underlying model without re-training \citep{jugovacjannachkarimi2018}.

\paragraph{Memory Requirements}
In terms of the memory requirements, the rule-based methods \ar, \mc, and \sr that basically record item co-occurrences of size two require the least memory, i.e., below 100 megabytes. Also the memory demands of \gru are very low in this comparison, and \gru occupies only about 60 MB of memory on the graphics card for the \emph{RSC15-S} dataset. The factorization-based methods and the neighborhood methods, in contrast, have substantially higher memory requirements. The lookup data structures of the neighborhood-based methods, for example, in our implementation occupy about 6 GB of memory. When additional recency-based sampling is applied, which according to the analyses above does not hurt accuracy, these demands could, however, be substantially lowered.

For some algorithms, the memory requirements largely depend on the characteristics of the datasets. Looking at the numbers for the \emph{8TRACKS} dataset, which covers over 300,000 different items (in contrast to the about 30,000 of the \emph{RSC15} dataset), we see that in particular the memory demand of \gru substantially increases with the number of items. As a result, \gru's network even needs more memory than neighborhood-based methods for this dataset. Given these observations it seems promising to implement additional data sampling strategies within the more complex methods---as we did for the nearest neighbor methods---to decrease their computational demands.

\section{Conclusion and Future Directions}
\label{sec:conclusions}
\subsection{Summary of Main Insights}
Being able to predict the user's short-term interest in an online session is a highly relevant problem in practice, which has raised increased interest also in the academic field in recent years. Even though a number of different algorithmic approaches were proposed over the years, no standard benchmark datasets and baseline algorithms exist today. In this work, we have compared a number of very recent and computationally complex algorithms for session-based recommendation with more light-weight approaches based, e.g., on session neighborhoods. The experimental analyses on a number of different datasets show that in many cases one of the simpler methods is able to outperform even the most recent methods based on recurrent neural networks in terms of the prediction accuracy. At the same time, the computational demands of these methods can be kept comparably low when using in-memory cache data structures and data sampling.

Overall, the results, therefore, indicate that additional research is required with respect to the development of sophisticated models that are more flexible in terms of how much sequential information is contained in the training data. This is in particular the case as several improvements for the nearest-neighbor methods can be imagined as well. In this work, we could for example observe that already using a different similarity measure, as done in the \vknn method, can lead to substantial performance improvements for different datasets. As a side result, we noticed that using the latent feature vectors of the items of the current session for sequential factorization-based methods does not lead to high accuracy values and that such methods are usually not strong baselines when comparing session-based algorithms.

Currently, constantly new deep learning-based algorithms for session-based recommendation are proposed, e.g., \citep{stamp2018} and \citep{binn2018}, which, for example, report improvements over \gru. We performed an initial evaluation of the STAMP method proposed in \citep{stamp2018}. Our first results indicate that STAMP does not outperform the trivial \sr technique in terms of the MRR on the \emph{Diginetica} dataset that was used for the evaluation in \citep{stamp2018}. The STAMP method, however, seems to be advantageous in terms of the hit rate for this particular dataset.

Generally, it is of course surprising that a recent and popular RNN-based method is not substantially better than longer-existing nearest neighbor approaches. We believe that this might be a result of the fact that for the specific task of session-based recommendation no ``standard'' existed so far with respect to baseline techniques and evaluation protocols. With this work, our aim is to contribute better baselines to benchmark session-based algorithms in the future.
A limitation of our work in some sense is that we could not identify \emph{one} best baseline method across all settings and datasets. While we would identify at least one very-well performing simpler method for each dataset, the relative performance of the algorithms seems depend on a number of factors, which are not yet fully understood.

Nevertheless, as the simple baseline approaches \ar, \sr, and \knn are computationally cheap and easy to test, their results obtained for a given dataset can potentially be used as an indicator for the general characteristics that a more complex model should aim to implement. If it, for example, turns out that \sr is the best performing baseline method, \gru, an extension to \gru or a different sequential model might probably be a good choice. In contrast, a good performance of \knn indicates that a more sophisticated model should not necessarily focus too much on the order of the items in a session.

\subsection{Future Directions}

From an algorithmic perspective, we believe that future complex models should consider more than the last event in a session when making the next-item prediction. Even in \gru, the previous items of a session are only considered implicitly through the hidden states in the prediction process. Our neighborhood models are in most cases much better when they consider all events in a session, albeit with a focus on the most recent interactions. In that context and in particular for longer sessions, it might also be helpful to detect interest changes that happen \emph{within} an individual session. This could, for example be achieved by considering \emph{semantic} information (e.g., meta-data or content features) about the items of the session, as was done, for example, in \citep{hariri12context} or \citep{Hidasi2016}. Recent advances in the area of deep learning might be particularly helpful in this context to extract such content features, e.g., from text, images, or videos, and to use this information in \emph{hybrid} approaches. Furthermore, the work in this paper focused on item-view events and more research is required to understand how to leverage other types of user actions like ``add-to-wishlist'' or ``add-to-cart'' in the learning and prediction process. With that information at hand, also other types of prediction problems can be addressed, e.g., whether or not a session will lead to a purchase or if there is a high probability that the user will abandon the session.

Going beyond the current session, more research also seems required in the area of \emph{session-aware} recommendation and the consideration of previous sessions of the current user. Open questions in this area are, for example, how to model general long-term user preferences (e.g., towards certain brands in e-commerce), how to detect user-individual preference drifts, or how to identify a subset of past sessions that are good predictors for the current one. This latter aspect was for example explored in \citep{Lerche2016} in the context of using recommendations as reminders. In addition, more elaborate strategies than static weighting schemes can be envisioned when combining short-term and long-term models. The importance weights, could for example be determined based on the length of the current session or the specific items that were considered.

Besides the consideration of signals at the individual user level, future research might also explore the incorporation of additional contextual factors or short-term trends in the community as a whole, when predicting the relevance of individual items. Recent works \citep{Tan2016,JannachLudewig2017RecSys, JannachLudewigLerche2017umuai} for example showed that considering short-term popularity trends and recency effects can lead to significant performance improvements in the e-commerce domain. Item recency (freshness) also plays a particular role in other domains such as music and news recommendation, and more work is required to understand how to integrate these aspects in today's recommendation algorithms.

Finally, since the relative performance of the different algorithms tested in our work sometimes varies across different datasets, more research is required to understand in which situations certain algorithms are better suited than others. These insights can then be further used to inform the design of hybrid recommendation approaches, which have shown to lead to the highest recommendation accuracy for session-based recommendation also in \citep{JannachLudewig2017RecSys}. Generally, many factors can influence the performance of a certain recommendation algorithm. \cite{Adomavicius:2012:IDC:2151163.2151166} have, for example, made a number of important analyses aiming to relate dataset characteristics, e.g., rating distributions and dataset sizes, with prediction accuracy. In the context of session-based recommendation problems, additional factors may have an influence, for example, the existence and strength of the sequential patterns that can be found in the data. Furthermore, often domain-specific aspects like item freshness and general item popularity might play important roles and should be further explored in future research.

\bibliographystyle{ACM-Reference-Format}
\bibliography{lit}

\newpage
\section*{Author Biographies}
\noindent \emph{Malte Ludewig} is a PhD candidate in Computer Science at TU Dortmund, Germany, from where he also received his MSc degree. His research interests lie in the field of recommender systems---with a focus on session-based recommendations---and personalization in e-commerce environments in general.

\vspace{6pt}
\noindent \emph{Dietmar Jannach} is a Professor of Computer Science at AAU Klagenfurt, Austria, and head of the department's information systems research group. Dr. Jannach has worked on different areas of artificial intelligence, including recommender systems, model-based diagnosis, and knowledge-based systems. He is the leading author of a textbook on recommender systems and has authored more than hundred research papers, focusing on the application of artificial intelligence technology to practical problems.

\newpage
\appendix

\section{Parameter Configurations}
\label{appx:params}

\begin{table}[H]
\scriptsize
\centering
\caption{Parameters for algorithm \gru for all datasets.}
\label{tab:appendix_window_gru_params}
\begin{tabular}{@{}l r r r r r @{}}
\toprule 
Dataset& layer\_size& objective& lr& momentum& drop\_out\\ 
\midrule 
RSC15 & 100 & $BPR_{max}$ & 0.20 & 0.5 & 0.0\\ 
TMALL & 100 & $TOP1_{max}$ & 0.05 & 0.0 & 0.3\\ 
RETAILROCKET & 100 & $TOP1_{max}$ & 0.15 & 0.3 & 0.0\\ 
8TRACKS & 100 & $TOP1_{max}$ & 0.10 & 0.0 & 0.7\\ 
AOTM & 100 & $BPR_{max}$ & 0.10 & 0.3 & 0.1\\ 
NOWPLAYING & 100 & $BPR_{max}$ & 0.10 & 0.5 & 0.1\\ 
30MUSIC & 100 & $TOP1_{max}$ & 0.10 & 0.1 & 0.1\\ 
ZALANDO & 100 & $BPR_{max}$ & 0.20 & 0.1 & 0.1\\ 
CLEF & 100 & $TOP1_{max}$ & 0.20 & 0.2 & 0.5\\ 
LASTFM & 100 & $BPR_{max}$ & 0.15 & 0.4 & 0.3\\ 
\bottomrule 
\end{tabular}
\end{table}
\begin{table}[H]
\scriptsize
\centering
\caption{Parameters used for the \smf algorithm for all datasets.}
\label{tab:appendix_window_smf_params}
\begin{tabular}{@{}l r r r r r r @{}}
\toprule
Dataset& layer\_size& objective& lr& momentum& drop\_out& skip\\
\midrule
RSC15 & 100 & $TOP1_{max}$ & 0.085 & 0.2 & 0.30 & 0.00\\
TMALL & 100 & $BPR_{max}$ & 0.015 & 0.6 & 0.00 & 0.00\\
RETAILROCKET & 100 & $BPR_{max}$ & 0.045 & 0.1 & 0.40 & 0.20\\
8TRACKS & 100 & $TOP1_{max}$ & 0.010 & 0.5 & 0.30 & 0.35\\
AOTM & 100 & $BPR_{max}$ & 0.090 & 0.8 & 0.40 & 0.20\\
NOWPLAYING & 100 & $TOP1_{max}$ & 0.055 & 0.2 & 0.40 & 0.20\\
30MUSIC & 100 & $TOP1_{max}$ & 0.100 & 0.1 & 0.20 & 0.20\\
ZALANDO & 100 & $TOP1_{max}$ & 0.030 & 0.3 & 0.25 & 0.00\\
CLEF & 100 & $BPR_{max}$ & 0.050 & 0.0 & 0.40 & 0.25\\
LASTFM & 100 & $TOP1_{max}$ & 0.015 & 0.3 & 0.15 & 0.45\\
\bottomrule
\end{tabular}
\end{table}
\input{tables_window_tex_params_vsknn.tex}
\begin{minipage}[t]{0.45\linewidth}
\scriptsize
\centering
\caption{Parameters used for the \sknn, \ssknn, and \sfknn algorithm for all datasets.}
\label{tab:appendix_window_sknn_params}
\begin{tabular}{@{}l r r @{}}
\toprule
Dataset& K& samples\\
\midrule
RSC15 & 500 & 1000\\
TMALL & 100 & 500\\
RETAILROCKET & 100 & 500\\
ZALANDO & 100 & 500\\
8TRACKS & 100 & 500\\
AOTM & 100 & 500\\
NOWPLAYING & 100 & 500\\
30MUSIC & 100 & 500\\
CLEF & 100 & 500\\
LASTFM & 100 & 500\\
\bottomrule
\end{tabular}
\end{minipage}
\end{table}

\pagebreak

\sisetup{
	round-mode          = places, 
	round-precision     = 3, 
}
\section{Full Result Tables}
\label{appx:full_results}

\begin{table}[H]
\tiny
\centering
\caption{Hit rate (HR), Mean reciprocal rank (MRR), item coverage (COV), and average popularity (POP) results for a list length of 20, 10, 5, 3, and 1 on the RSC15 dataset (sorted by MRR@20).}
\label{tab:appendix_window_rsc15_results}
\begin{tabular}{@{}l r r r r r r r r @{}}
\toprule 
Algorithm& MRR@20& HR@20& COV@20& POP@20& MRR@10& HR@10& COV@10& POP@10\\ 
\midrule 
\gru & \num{0.307894} & \num{0.682701} & \num{0.504040} & \num{0.054114} & \num{0.301400} & \num{0.590551} & \num{0.430729} & \num{0.058044}\\ 
\sr & \num{0.303728} & \num{0.653089} & \num{0.667697} & \num{0.071734} & \num{0.297780} & \num{0.568726} & \num{0.592493} & \num{0.072593}\\ 
\smf & \num{0.301606} & \num{0.665859} & \num{0.565263} & \num{0.054524} & \num{0.295201} & \num{0.574904} & \num{0.485636} & \num{0.058009}\\ 
\mc & \num{0.300453} & \num{0.641768} & \num{0.645401} & \num{0.070062} & \num{0.294824} & \num{0.562030} & \num{0.584225} & \num{0.070626}\\ 
\ar & \num{0.289403} & \num{0.636098} & \num{0.629634} & \num{0.092633} & \num{0.283334} & \num{0.549753} & \num{0.548119} & \num{0.091246}\\ 
\vsknn & \num{0.282850} & \num{0.653048} & \num{0.618683} & \num{0.078584} & \num{0.276749} & \num{0.562769} & \num{0.534283} & \num{0.081186}\\ 
\ssknn & \num{0.271644} & \num{0.601905} & \num{0.654886} & \num{0.071569} & \num{0.266572} & \num{0.530621} & \num{0.543019} & \num{0.076820}\\
\bottomrule 
\end{tabular}
\end{table}
\begin{table}[H]
\tiny
\centering
\caption{Hit rate (HR), Mean reciprocal rank (MRR), item coverage (COV), and average popularity (POP) results for a list length of 20, 10, 5, 3, and 1 on the RSC15-S dataset (sorted by MRR@20).}
\label{tab:appendix_window_rsc15s_results}
\begin{tabular}{@{}l r r r r r r r r @{}}
\toprule 
Algorithm& MRR@20& HR@20& COV@20& POP@20& MRR@10& HR@10& COV@10& POP@10\\ 
\midrule 
\smf & \num{0.308673} & \num{0.712664} & \num{0.512132} & \num{0.051638} & \num{0.301127} & \num{0.605795} & \num{0.414102} & \num{0.053844}\\ 
\gru & \num{0.308486} & \num{0.718720} & \num{0.350452} & \num{0.032746} & \num{0.300723} & \num{0.608984} & \num{0.276605} & \num{0.034398}\\ 
\sr & \num{0.307549} & \num{0.690329} & \num{0.511645} & \num{0.038156} & \num{0.300557} & \num{0.590755} & \num{0.406985} & \num{0.037643}\\ 
\mc & \num{0.296144} & \num{0.666857} & \num{0.517781} & \num{0.039044} & \num{0.289091} & \num{0.566568} & \num{0.413387} & \num{0.038557}\\ 
\ar & \num{0.280786} & \num{0.655193} & \num{0.472881} & \num{0.044855} & \num{0.272944} & \num{0.543275} & \num{0.374436} & \num{0.042533}\\ 
\vsknn & \num{0.273589} & \num{0.675212} & \num{0.427180} & \num{0.036745} & \num{0.265621} & \num{0.562328} & \num{0.327695} & \num{0.038544}\\ 
\ssknn & \num{0.266495} & \num{0.666607} & \num{0.417336} & \num{0.035487} & \num{0.258178} & \num{0.548481} & \num{0.308673} & \num{0.037991}\\
\bottomrule 
\end{tabular}
\end{table}
\begin{table}[H]
\tiny
\centering
\caption{Hit rate (HR), Mean reciprocal rank (MRR), item coverage (COV), and average popularity (POP) results for a list length of 20, 10, 5, 3, and 1 on the TMALL dataset (sorted by MRR@20).}
\label{tab:appendix_window_tmall_results}
\begin{tabular}{@{}l r r r r r r r r @{}}
\toprule 
Algorithm& MRR@20& HR@20& COV@20& POP@20& MRR@10& HR@10& COV@10& POP@10\\ 
\midrule 
\ssknn & \num{0.185263} & \num{0.386548} & \num{0.466752} & \num{0.024519} & \num{0.181325} & \num{0.330241} & \num{0.308861} & \num{0.027423}\\ 
\knn & \num{0.181547} & \num{0.403854} & \num{0.380797} & \num{0.026072} & \num{0.177044} & \num{0.334347} & \num{0.248608} & \num{0.028883}\\ 
\vsknn & \num{0.178912} & \num{0.373288} & \num{0.464475} & \num{0.023561} & \num{0.174648} & \num{0.311911} & \num{0.319940} & \num{0.026078}\\ 
\bpr & \num{0.159065} & \num{0.204018} & \num{0.722647} & \num{0.057255} & \num{0.158565} & \num{0.196798} & \num{0.534302} & \num{0.076121}\\ 
\sfsknn & \num{0.135712} & \num{0.216377} & \num{0.435876} & \num{0.018155} & \num{0.134787} & \num{0.203465} & \num{0.338131} & \num{0.022173}\\ 
\gru & \num{0.129059} & \num{0.277065} & \num{0.151262} & \num{0.035479} & \num{0.125458} & \num{0.224967} & \num{0.108674} & \num{0.039124}\\ 
\ar & \num{0.128903} & \num{0.262101} & \num{0.509116} & \num{0.021389} & \num{0.125795} & \num{0.217170} & \num{0.357970} & \num{0.023768}\\
\bottomrule 
\end{tabular}
\end{table}
\begin{table}[H]
\tiny
\centering
\caption{Hit rate (HR), Mean reciprocal rank (MRR), item coverage (COV), and average popularity (POP) results for a list length of 20, 10, 5, 3, and 1 on the RETAILROCKET dataset (sorted by MRR@20).}
\label{tab:appendix_window_retailrocket_results}
\begin{tabular}{@{}l r r r r r r r r @{}}
\toprule 
Algorithm& MRR@20& HR@20& COV@20& POP@20& MRR@10& HR@10& COV@10& POP@10\\ 
\midrule 
\ssknn & \num{0.345067} & \num{0.590670} & \num{0.595784} & \num{0.056183} & \num{0.341314} & \num{0.537347} & \num{0.479856} & \num{0.065730}\\ 
\vsknn & \num{0.337884} & \num{0.572673} & \num{0.575407} & \num{0.059880} & \num{0.334172} & \num{0.519353} & \num{0.473680} & \num{0.069153}\\ 
\knn & \num{0.337125} & \num{0.583010} & \num{0.565694} & \num{0.057723} & \num{0.333304} & \num{0.528440} & \num{0.445218} & \num{0.068234}\\ 
\bpr & \num{0.302970} & \num{0.356996} & \num{0.824149} & \num{0.060233} & \num{0.302598} & \num{0.351712} & \num{0.627335} & \num{0.071787}\\ 
\fpmc & \num{0.272940} & \num{0.320056} & \num{0.929133} & \num{0.022468} & \num{0.272166} & \num{0.308743} & \num{0.777176} & \num{0.026088}\\ 
\sfsknn & \num{0.259962} & \num{0.357631} & \num{0.403284} & \num{0.034541} & \num{0.259426} & \num{0.350478} & \num{0.373038} & \num{0.046101}\\ 
\sr & \num{0.245440} & \num{0.418655} & \num{0.524053} & \num{0.041945} & \num{0.243121} & \num{0.386228} & \num{0.458418} & \num{0.050438}\\
\bottomrule 
\end{tabular}
\end{table}
\begin{table}[H]
\tiny
\centering
\caption{Hit rate (HR), Mean reciprocal rank (MRR), item coverage (COV), and average popularity (POP) results for a list length of 20, 10, 5, 3, and 1 on the ZALANDO dataset (sorted by MRR@20).}
\label{tab:appendix_window_zalando_results}
\begin{tabular}{@{}l r r r r r r r r @{}}
\toprule 
Algorithm& MRR@20& HR@20& COV@20& POP@20& MRR@10& HR@10& COV@10& POP@10\\ 
\midrule 
\sr & \num{0.303819} & \num{0.482965} & \num{0.586016} & \num{0.061204} & \num{0.302333} & \num{0.461891} & \num{0.433177} & \num{0.065634}\\ 
\mc & \num{0.302646} & \num{0.455389} & \num{0.512848} & \num{0.059965} & \num{0.301622} & \num{0.440988} & \num{0.411576} & \num{0.066272}\\ 
\iknn & \num{0.274582} & \num{0.404771} & \num{0.714105} & \num{0.036944} & \num{0.273245} & \num{0.385450} & \num{0.532421} & \num{0.041177}\\ 
\gru & \num{0.267160} & \num{0.467861} & \num{0.303882} & \num{0.100605} & \num{0.264744} & \num{0.433131} & \num{0.238626} & \num{0.103022}\\ 
\smf & \num{0.266810} & \num{0.447025} & \num{0.361635} & \num{0.107216} & \num{0.264800} & \num{0.418241} & \num{0.281748} & \num{0.107573}\\ 
\ar & \num{0.257847} & \num{0.466512} & \num{0.467207} & \num{0.088640} & \num{0.255632} & \num{0.434547} & \num{0.336746} & \num{0.090210}\\ 
\sfsknn & \num{0.249494} & \num{0.437780} & \num{0.432409} & \num{0.056822} & \num{0.248903} & \num{0.429667} & \num{0.348298} & \num{0.068323}\\
\bottomrule 
\end{tabular}
\end{table}

\sisetup{
	round-mode          = places, 
	round-precision     = 4, 
}

\begin{table}[H]
\tiny
\centering
\caption{Hit rate (HR), Mean reciprocal rank (MRR), item coverage (COV), and average popularity (POP) results for a list length of 20, 10, 5, 3, and 1 on the 8TRACKS dataset (sorted by MRR@20).}
\label{tab:appendix_window_8tracks_results}
\begin{tabular}{@{}l r r r r r r r r @{}}
\toprule 
Algorithm& MRR@20& HR@20& COV@20& POP@20& MRR@10& HR@10& COV@10& POP@10\\ 
\midrule 
\ar & \num{0.007134} & \num{0.025537} & \num{0.452959} & \num{0.091245} & \num{0.006573} & \num{0.017338} & \num{0.317815} & \num{0.107481}\\ 
\smf & \num{0.006411} & \num{0.023097} & \num{0.152798} & \num{0.086446} & \num{0.005848} & \num{0.014831} & \num{0.107573} & \num{0.091620}\\ 
\sr & \num{0.006388} & \num{0.017078} & \num{0.496712} & \num{0.053106} & \num{0.006077} & \num{0.012543} & \num{0.364503} & \num{0.063564}\\ 
\sfsknn & \num{0.006373} & \num{0.011876} & \num{0.304936} & \num{0.036284} & \num{0.006255} & \num{0.010158} & \num{0.243875} & \num{0.051668}\\ 
\vsknn & \num{0.005729} & \num{0.035221} & \num{0.408058} & \num{0.119431} & \num{0.004767} & \num{0.021119} & \num{0.252304} & \num{0.135259}\\ 
\knn & \num{0.005347} & \num{0.037553} & \num{0.243081} & \num{0.107967} & \num{0.004133} & \num{0.019761} & \num{0.154369} & \num{0.115286}\\ 
\iknn & \num{0.005058} & \num{0.017672} & \num{0.695607} & \num{0.024539} & \num{0.004694} & \num{0.012376} & \num{0.510125} & \num{0.026666}\\
\bottomrule 
\end{tabular}
\end{table}
\begin{table}[H]
\tiny
\centering
\caption{Hit rate (HR), Mean reciprocal rank (MRR), item coverage (COV), and average popularity (POP) results for a list length of 20, 10, 5, 3, and 1 on the AOTM dataset (sorted by MRR@20).}
\label{tab:appendix_window_aotm_results}
\begin{tabular}{@{}l r r r r r r r r @{}}
\toprule 
Algorithm& MRR@20& HR@20& COV@20& POP@20& MRR@10& HR@10& COV@10& POP@10\\ 
\midrule 
\smf & \num{0.011114} & \num{0.029780} & \num{0.245659} & \num{0.199819} & \num{0.010479} & \num{0.020465} & \num{0.179468} & \num{0.208476}\\ 
\sfsknn & \num{0.011067} & \num{0.014480} & \num{0.355883} & \num{0.050836} & \num{0.011027} & \num{0.013904} & \num{0.302223} & \num{0.068554}\\ 
\sr & \num{0.007662} & \num{0.019549} & \num{0.586409} & \num{0.053319} & \num{0.007345} & \num{0.014883} & \num{0.448115} & \num{0.059920}\\ 
\gru & \num{0.007185} & \num{0.015661} & \num{0.465274} & \num{0.115128} & \num{0.006967} & \num{0.012480} & \num{0.354963} & \num{0.113075}\\ 
\mc & \num{0.006321} & \num{0.013284} & \num{0.380324} & \num{0.049764} & \num{0.006178} & \num{0.011189} & \num{0.319832} & \num{0.060160}\\ 
\ar & \num{0.005918} & \num{0.023329} & \num{0.553178} & \num{0.104921} & \num{0.005334} & \num{0.014629} & \num{0.400297} & \num{0.117849}\\ 
\vsknn & \num{0.005488} & \num{0.037781} & \num{0.536266} & \num{0.139713} & \num{0.004326} & \num{0.020771} & \num{0.335697} & \num{0.155048}\\
\bottomrule 
\end{tabular}
\end{table}
\begin{table}[H]
\tiny
\centering
\caption{Hit rate (HR), Mean reciprocal rank (MRR), item coverage (COV), and average popularity (POP) results for a list length of 20, 10, 5, 3, and 1 on the 30MUSIC dataset (sorted by MRR@20).}
\label{tab:appendix_window_30music_results}
\begin{tabular}{@{}l r r r r r r r r @{}}
\toprule 
Algorithm& MRR@20& HR@20& COV@20& POP@20& MRR@10& HR@10& COV@10& POP@10\\ 
\midrule 
\sr & \num{0.237721} & \num{0.332310} & \num{0.389255} & \num{0.023192} & \num{0.236291} & \num{0.312018} & \num{0.291341} & \num{0.027327}\\ 
\mc & \num{0.231813} & \num{0.284376} & \num{0.203840} & \num{0.020522} & \num{0.231358} & \num{0.277952} & \num{0.180354} & \num{0.026464}\\ 
\gru & \num{0.226371} & \num{0.325692} & \num{0.344739} & \num{0.055627} & \num{0.224867} & \num{0.304176} & \num{0.242252} & \num{0.056653}\\ 
\sfsknn & \num{0.207934} & \num{0.285608} & \num{0.185429} & \num{0.021948} & \num{0.207763} & \num{0.283350} & \num{0.163416} & \num{0.030229}\\ 
\smf & \num{0.177651} & \num{0.284285} & \num{0.150848} & \num{0.104775} & \num{0.175576} & \num{0.254672} & \num{0.111702} & \num{0.106196}\\ 
\vsknn & \num{0.109873} & \num{0.381867} & \num{0.316960} & \num{0.053790} & \num{0.104042} & \num{0.300152} & \num{0.194441} & \num{0.057256}\\ 
\iknn & \num{0.108610} & \num{0.297108} & \num{0.459578} & \num{0.022591} & \num{0.105252} & \num{0.250115} & \num{0.312240} & \num{0.024932}\\
\bottomrule 
\end{tabular}
\end{table}
\begin{table}[H]
\tiny
\centering
\caption{Hit rate (HR), Mean reciprocal rank (MRR), item coverage (COV), and average popularity (POP) results for a list length of 20, 10, 5, 3, and 1 on the NOWPLAYING dataset (sorted by MRR@20).}
\label{tab:appendix_window_nowplaying_results}
\begin{tabular}{@{}l r r r r r r r r @{}}
\toprule 
Algorithm& MRR@20& HR@20& COV@20& POP@20& MRR@10& HR@10& COV@10& POP@10\\ 
\midrule 
\sr & \num{0.105282} & \num{0.203301} & \num{0.465555} & \num{0.024730} & \num{0.103054} & \num{0.171183} & \num{0.360491} & \num{0.028417}\\ 
\gru & \num{0.101838} & \num{0.196971} & \num{0.433107} & \num{0.051596} & \num{0.099522} & \num{0.163157} & \num{0.326092} & \num{0.052865}\\ 
\mc & \num{0.097104} & \num{0.158193} & \num{0.293560} & \num{0.028392} & \num{0.095950} & \num{0.141736} & \num{0.254721} & \num{0.034738}\\ 
\sfsknn & \num{0.095425} & \num{0.164721} & \num{0.277271} & \num{0.031125} & \num{0.094542} & \num{0.152353} & \num{0.236868} & \num{0.041408}\\ 
\smf & \num{0.088197} & \num{0.182504} & \num{0.241696} & \num{0.091586} & \num{0.085860} & \num{0.148353} & \num{0.184716} & \num{0.096008}\\ 
\vsknn & \num{0.078463} & \num{0.255195} & \num{0.428299} & \num{0.063943} & \num{0.073657} & \num{0.186080} & \num{0.290430} & \num{0.071916}\\ 
\ssknn & \num{0.077679} & \num{0.262175} & \num{0.414882} & \num{0.062390} & \num{0.072528} & \num{0.188033} & \num{0.272654} & \num{0.069922}\\
\bottomrule 
\end{tabular}
\end{table}

\sisetup{
	round-mode          = places, 
	round-precision     = 3, 
}

\begin{table}[H]
\tiny
\centering
\caption{Hit rate (HR), Mean reciprocal rank (MRR), item coverage (COV), and average popularity (POP) results for a list length of 20, 10, 5, 3, and 1 on the CLEF dataset (sorted by MRR@20).}
\label{tab:appendix_window_clef_results}
\begin{tabular}{@{}l r r r r r r r r @{}}
\toprule 
Algorithm& MRR@20& HR@20& COV@20& POP@20& MRR@10& HR@10& COV@10& POP@10\\ 
\midrule 
\smf & \num{0.234424} & \num{0.706117} & \num{0.649565} & \num{0.082843} & \num{0.222067} & \num{0.529420} & \num{0.581601} & \num{0.096621}\\ 
\mc & \num{0.225302} & \num{0.687346} & \num{0.731665} & \num{0.094838} & \num{0.213319} & \num{0.514217} & \num{0.705066} & \num{0.123164}\\ 
\vsknn & \num{0.223575} & \num{0.775558} & \num{0.621024} & \num{0.082179} & \num{0.211030} & \num{0.596319} & \num{0.565946} & \num{0.113202}\\ 
\sr & \num{0.222966} & \num{0.671878} & \num{0.655356} & \num{0.093288} & \num{0.211955} & \num{0.512647} & \num{0.607898} & \num{0.123337}\\ 
\gru & \num{0.219950} & \num{0.568499} & \num{0.174425} & \num{0.094286} & \num{0.212333} & \num{0.462381} & \num{0.129061} & \num{0.117711}\\ 
\knn & \num{0.218566} & \num{0.778078} & \num{0.612573} & \num{0.084236} & \num{0.205215} & \num{0.587774} & \num{0.544552} & \num{0.122057}\\ 
\ar & \num{0.216195} & \num{0.665970} & \num{0.723702} & \num{0.099540} & \num{0.203563} & \num{0.489993} & \num{0.656406} & \num{0.147938}\\
\bottomrule 
\end{tabular}
\end{table}

\section{Additional Results for Precision and Recall}
\label{appx:pr_results}

\begin{table}[H]
\tiny
\centering
\caption{Precision (P) and Recall (R) results for a list length of 20, 10, 5, and 3 on the TMALL dataset (sorted by P@20).}
\label{tab:appendix_pr_tmall_results}
\begin{tabular}{@{}l r r r r r r r r @{}}
\toprule 
Algorithm& P@20& R@20& P@10& R@10& P@5& R@5& P@3& R@3\\ 
\midrule 
\knn & \num{0.094525} & \num{0.312230} & \num{0.140941} & \num{0.256922} & \num{0.196174} & \num{0.198951} & \num{0.235196} & \num{0.156265}\\ 
\ssknn & \num{0.093886} & \num{0.262890} & \num{0.138721} & \num{0.214856} & \num{0.191017} & \num{0.164720} & \num{0.228788} & \num{0.128710}\\ 
\vsknn & \num{0.090787} & \num{0.290696} & \num{0.131043} & \num{0.238816} & \num{0.186093} & \num{0.186995} & \num{0.229772} & \num{0.149905}\\ 
\smf & \num{0.067891} & \num{0.230440} & \num{0.098986} & \num{0.184027} & \num{0.138536} & \num{0.141369} & \num{0.172280} & \num{0.112623}\\ 
\gru & \num{0.067669} & \num{0.233084} & \num{0.098253} & \num{0.186636} & \num{0.137143} & \num{0.143066} & \num{0.170233} & \num{0.114599}\\ 
\ar & \num{0.056701} & \num{0.173098} & \num{0.082042} & \num{0.137826} & \num{0.114969} & \num{0.105979} & \num{0.143178} & \num{0.084617}\\ 
\sr & \num{0.051922} & \num{0.192828} & \num{0.080952} & \num{0.161627} & \num{0.121090} & \num{0.130561} & \num{0.158452} & \num{0.108776}\\
\bottomrule 
\end{tabular}
\end{table}
\begin{table}[H]
\tiny
\centering
\caption{Precision (P) and Recall (R) results for a list length of 20, 10, 5, and 3 on the RETAILROCKET dataset (sorted by P@20).}
\label{tab:appendix_pr_retailrocket_results}
\begin{tabular}{@{}l r r r r r r r r @{}}
\toprule 
Algorithm& P@20& R@20& P@10& R@10& P@5& R@5& P@3& R@3\\ 
\midrule 
\ssknn & \num{0.056879} & \num{0.479760} & \num{0.095543} & \num{0.433595} & \num{0.155528} & \num{0.375936} & \num{0.214353} & \num{0.326467}\\ 
\knn & \num{0.056196} & \num{0.477884} & \num{0.094758} & \num{0.432674} & \num{0.152001} & \num{0.373028} & \num{0.207926} & \num{0.322918}\\ 
\vsknn & \num{0.055266} & \num{0.462154} & \num{0.092966} & \num{0.417068} & \num{0.151613} & \num{0.363843} & \num{0.209024} & \num{0.316137}\\ 
\smf & \num{0.047137} & \num{0.397138} & \num{0.073627} & \num{0.334885} & \num{0.112935} & \num{0.270684} & \num{0.147660} & \num{0.218875}\\ 
\gru & \num{0.045819} & \num{0.399711} & \num{0.073006} & \num{0.339224} & \num{0.111462} & \num{0.270532} & \num{0.143365} & \num{0.214799}\\ 
\ar & \num{0.040762} & \num{0.360140} & \num{0.068084} & \num{0.318124} & \num{0.109389} & \num{0.267624} & \num{0.146885} & \num{0.224728}\\ 
\sr & \num{0.038039} & \num{0.341720} & \num{0.066825} & \num{0.312813} & \num{0.109621} & \num{0.268932} & \num{0.149243} & \num{0.227081}\\
\bottomrule 
\end{tabular}
\end{table}
\begin{table}[H]
\tiny
\centering
\caption{Precision (P) and Recall (R) results for a list length of 20, 10, 5, and 3 on the ZALANDO dataset (sorted by P@20).}
\label{tab:appendix_pr_zalando_results}
\begin{tabular}{@{}l r r r r r r r r @{}}
\toprule 
Algorithm& P@20& R@20& P@10& R@10& P@5& R@5& P@3& R@3\\ 
\midrule 
\vsknn & \num{0.075748} & \num{0.218740} & \num{0.129373} & \num{0.195308} & \num{0.210409} & \num{0.167844} & \num{0.282197} & \num{0.140724}\\ 
\ssknn & \num{0.074583} & \num{0.217329} & \num{0.126536} & \num{0.192322} & \num{0.198279} & \num{0.158664} & \num{0.252435} & \num{0.126762}\\ 
\knn & \num{0.074167} & \num{0.201776} & \num{0.114583} & \num{0.168885} & \num{0.163888} & \num{0.129569} & \num{0.195889} & \num{0.098440}\\ 
\gru & \num{0.064554} & \num{0.181378} & \num{0.108707} & \num{0.161004} & \num{0.181322} & \num{0.140950} & \num{0.246865} & \num{0.119575}\\ 
\smf & \num{0.061615} & \num{0.175186} & \num{0.102808} & \num{0.154147} & \num{0.169115} & \num{0.133486} & \num{0.227785} & \num{0.112344}\\ 
\ar & \num{0.060234} & \num{0.179407} & \num{0.105309} & \num{0.161473} & \num{0.177539} & \num{0.141024} & \num{0.240612} & \num{0.118099}\\ 
\sr & \num{0.059531} & \num{0.174376} & \num{0.106171} & \num{0.161029} & \num{0.185240} & \num{0.145628} & \num{0.260181} & \num{0.127477}\\
\bottomrule 
\end{tabular}
\end{table}

\sisetup{
	round-mode          = places, 
	round-precision     = 4, 
}

\begin{table}[H]
\tiny
\centering
\caption{Precision (P) and Recall (R) results for a list length of 20, 10, 5, and 3 on the 8TRACKS dataset (sorted by P@20).}
\label{tab:appendix_pr_8tracks_results}
\begin{tabular}{@{}l r r r r r r r r @{}}
\toprule 
Algorithm& P@20& R@20& P@10& R@10& P@5& R@5& P@3& R@3\\ 
\midrule 
\vsknn & \num{0.012174} & \num{0.030801} & \num{0.013802} & \num{0.018429} & \num{0.012479} & \num{0.008427} & \num{0.010894} & \num{0.004658}\\ 
\knn & \num{0.011737} & \num{0.031319} & \num{0.011905} & \num{0.017099} & \num{0.009009} & \num{0.006735} & \num{0.005855} & \num{0.002590}\\ 
\ssknn & \num{0.010015} & \num{0.026588} & \num{0.011395} & \num{0.016179} & \num{0.009707} & \num{0.006997} & \num{0.006543} & \num{0.002873}\\ 
\ar & \num{0.008747} & \num{0.021911} & \num{0.011230} & \num{0.014669} & \num{0.013593} & \num{0.009019} & \num{0.015535} & \num{0.006221}\\ 
\smf & \num{0.008609} & \num{0.021782} & \num{0.010383} & \num{0.013463} & \num{0.011757} & \num{0.007806} & \num{0.012627} & \num{0.005190}\\ 
\iknn & \num{0.006025} & \num{0.014899} & \num{0.007746} & \num{0.010234} & \num{0.009160} & \num{0.006250} & \num{0.010307} & \num{0.004335}\\ 
\sr & \num{0.005519} & \num{0.013996} & \num{0.007749} & \num{0.009891} & \num{0.009845} & \num{0.006674} & \num{0.011101} & \num{0.004607}\\
\bottomrule 
\end{tabular}
\end{table}
\begin{table}[H]
\tiny
\centering
\caption{Precision (P) and Recall (R) results for a list length of 20, 10, 5, and 3 on the AOTM dataset (sorted by P@20).}
\label{tab:appendix_pr_aotm_results}
\begin{tabular}{@{}l r r r r r r r r @{}}
\toprule 
Algorithm& P@20& R@20& P@10& R@10& P@5& R@5& P@3& R@3\\ 
\midrule 
\knn & \num{0.015502} & \num{0.044026} & \num{0.015744} & \num{0.021403} & \num{0.009897} & \num{0.005803} & \num{0.005557} & \num{0.002311}\\ 
\vsknn & \num{0.013318} & \num{0.036128} & \num{0.014492} & \num{0.019581} & \num{0.011829} & \num{0.007788} & \num{0.010618} & \num{0.005642}\\ 
\ssknn & \num{0.012481} & \num{0.035312} & \num{0.012235} & \num{0.017777} & \num{0.008428} & \num{0.006507} & \num{0.005373} & \num{0.002704}\\ 
\smf & \num{0.008400} & \num{0.025856} & \num{0.010508} & \num{0.016287} & \num{0.012960} & \num{0.010443} & \num{0.014333} & \num{0.007033}\\ 
\ar & \num{0.006563} & \num{0.018325} & \num{0.008244} & \num{0.011876} & \num{0.009549} & \num{0.006803} & \num{0.010720} & \num{0.004854}\\ 
\iknn & \num{0.005569} & \num{0.015532} & \num{0.006893} & \num{0.010200} & \num{0.008213} & \num{0.006233} & \num{0.009006} & \num{0.004336}\\ 
\sr & \num{0.005345} & \num{0.014590} & \num{0.006963} & \num{0.009759} & \num{0.008223} & \num{0.006069} & \num{0.009153} & \num{0.004109}\\
\bottomrule 
\end{tabular}
\end{table}
\begin{table}[H]
\tiny
\centering
\caption{Precision (P) and Recall (R) results for a list length of 20, 10, 5, and 3 on the 30MUSIC dataset (sorted by P@20).}
\label{tab:appendix_pr_30music_results}
\begin{tabular}{@{}l r r r r r r r r @{}}
\toprule 
Algorithm& P@20& R@20& P@10& R@10& P@5& R@5& P@3& R@3\\ 
\midrule 
\vsknn & \num{0.111670} & \num{0.243838} & \num{0.158520} & \num{0.194792} & \num{0.194709} & \num{0.137119} & \num{0.223925} & \num{0.104444}\\ 
\ssknn & \num{0.111011} & \num{0.235272} & \num{0.143911} & \num{0.167172} & \num{0.143131} & \num{0.083199} & \num{0.134449} & \num{0.045756}\\ 
\knn & \num{0.103533} & \num{0.213986} & \num{0.129548} & \num{0.146183} & \num{0.128289} & \num{0.072173} & \num{0.121587} & \num{0.040232}\\ 
\iknn & \num{0.093516} & \num{0.202283} & \num{0.133607} & \num{0.161072} & \num{0.152901} & \num{0.101545} & \num{0.158502} & \num{0.065117}\\ 
\ar & \num{0.091439} & \num{0.192270} & \num{0.124375} & \num{0.143484} & \num{0.135404} & \num{0.082466} & \num{0.138585} & \num{0.051400}\\ 
\sr & \num{0.087771} & \num{0.200965} & \num{0.139299} & \num{0.175004} & \num{0.188360} & \num{0.135274} & \num{0.220352} & \num{0.104165}\\ 
\smf & \num{0.074615} & \num{0.165541} & \num{0.102490} & \num{0.127166} & \num{0.128982} & \num{0.087583} & \num{0.145062} & \num{0.062641}\\
\bottomrule 
\end{tabular}
\end{table}
\begin{table}[H]
\tiny
\centering
\caption{Precision (P) and Recall (R) results for a list length of 20, 10, 5, and 3 on the NOWPLAYING dataset (sorted by P@20).}
\label{tab:appendix_pr_nowplaying_results}
\begin{tabular}{@{}l r r r r r r r r @{}}
\toprule 
Algorithm& P@20& R@20& P@10& R@10& P@5& R@5& P@3& R@3\\ 
\midrule 
\ssknn & \num{0.072616} & \num{0.194392} & \num{0.089021} & \num{0.129598} & \num{0.094981} & \num{0.069444} & \num{0.093486} & \num{0.040471}\\ 
\vsknn & \num{0.071826} & \num{0.190934} & \num{0.089992} & \num{0.130281} & \num{0.104839} & \num{0.087297} & \num{0.116864} & \num{0.063317}\\ 
\knn & \num{0.068007} & \num{0.182355} & \num{0.084107} & \num{0.118553} & \num{0.086809} & \num{0.062225} & \num{0.088996} & \num{0.036180}\\ 
\ar & \num{0.055365} & \num{0.155076} & \num{0.072436} & \num{0.108637} & \num{0.087576} & \num{0.070535} & \num{0.096847} & \num{0.049114}\\ 
\sr & \num{0.050113} & \num{0.146467} & \num{0.071728} & \num{0.113243} & \num{0.094499} & \num{0.082550} & \num{0.107984} & \num{0.061406}\\ 
\smf & \num{0.049865} & \num{0.145303} & \num{0.066785} & \num{0.104298} & \num{0.084039} & \num{0.070940} & \num{0.096621} & \num{0.052532}\\ 
\iknn & \num{0.049206} & \num{0.138469} & \num{0.063910} & \num{0.097410} & \num{0.075536} & \num{0.060779} & \num{0.080894} & \num{0.040480}\\
\bottomrule 
\end{tabular}
\end{table}

\sisetup{
	round-mode          = places, 
	round-precision     = 3, 
}

\begin{table}[H]
\tiny
\centering
\caption{Precision (P) and Recall (R) results for a list length of 20, 10, 5, and 3 on the CLEF dataset (sorted by P@20).}
\label{tab:appendix_pr_clef_results}
\begin{tabular}{@{}l r r r r r r r r @{}}
\toprule 
Algorithm& P@20& R@20& P@10& R@10& P@5& R@5& P@3& R@3\\ 
\midrule 
\gru & \num{0.072448} & \num{0.626009} & \num{0.100185} & \num{0.453503} & \num{0.128179} & \num{0.297575} & \num{0.144321} & \num{0.204325}\\ 
\vsknn & \num{0.068508} & \num{0.593304} & \num{0.089087} & \num{0.412883} & \num{0.108178} & \num{0.261585} & \num{0.118589} & \num{0.179789}\\ 
\ssknn & \num{0.065862} & \num{0.579145} & \num{0.085911} & \num{0.404382} & \num{0.097056} & \num{0.244327} & \num{0.100711} & \num{0.154255}\\ 
\knn & \num{0.065616} & \num{0.577133} & \num{0.084614} & \num{0.399356} & \num{0.095640} & \num{0.241278} & \num{0.098844} & \num{0.152073}\\ 
\sfsknn & \num{0.063991} & \num{0.564560} & \num{0.082369} & \num{0.389946} & \num{0.094914} & \num{0.240522} & \num{0.098395} & \num{0.151476}\\ 
\smf & \num{0.061677} & \num{0.526821} & \num{0.084010} & \num{0.376897} & \num{0.106932} & \num{0.246217} & \num{0.119607} & \num{0.167022}\\ 
\fpmc & \num{0.060151} & \num{0.514585} & \num{0.078021} & \num{0.347236} & \num{0.093352} & \num{0.216430} & \num{0.109852} & \num{0.154343}\\
\bottomrule 
\end{tabular}
\end{table}


\section{Additional Single Split Results}
\label{appx:singleres}

\begin{table}[H]
\tiny
\centering
\caption{Hit rate (HR), Mean reciprocal rank (MRR), item coverage (COV), and average popularity (POP) results for a list length of 20, 10, 5, 3, and 1 on the TMALL dataset with a single split (sorted by MRR@20).}
\label{tab:appendix_window_tmall_results_single}
\begin{tabular}{@{}l r r r r r r r r @{}}
\toprule 
Algorithm& MRR@20& HR@20& COV@20& POP@20& MRR@10& HR@10& COV@10& POP@10\\ 
\midrule 
\ssknn & \num{0.180612} & \num{0.384580} & \num{0.559537} & \num{0.018911} & \num{0.176738} & \num{0.329361} & \num{0.374786} & \num{0.020898}\\ 
\knn & \num{0.177128} & \num{0.398333} & \num{0.461346} & \num{0.019891} & \num{0.172613} & \num{0.333937} & \num{0.305804} & \num{0.021806}\\ 
\vsknn & \num{0.169400} & \num{0.353256} & \num{0.569571} & \num{0.019141} & \num{0.165603} & \num{0.298480} & \num{0.404680} & \num{0.020838}\\ 
\sfsknn & \num{0.141239} & \num{0.237177} & \num{0.577260} & \num{0.015414} & \num{0.140029} & \num{0.220281} & \num{0.440944} & \num{0.018048}\\ 
\smf & \num{0.139764} & \num{0.297602} & \num{0.461013} & \num{0.023376} & \num{0.136121} & \num{0.244877} & \num{0.341089} & \num{0.023398}\\ 
\ar & \num{0.131210} & \num{0.254028} & \num{0.628133} & \num{0.018838} & \num{0.128440} & \num{0.214077} & \num{0.456624} & \num{0.020608}\\ 
\sr & \num{0.130906} & \num{0.242788} & \num{0.683263} & \num{0.018632} & \num{0.128528} & \num{0.208573} & \num{0.507069} & \num{0.019938}\\
\bottomrule 
\end{tabular}
\end{table}
\begin{table}[H]
\tiny
\centering
\caption{Hit rate (HR), Mean reciprocal rank (MRR), item coverage (COV), and average popularity (POP) results for a list length of 20, 10, 5, 3, and 1 on the RETAILROCKET dataset with a single split (sorted by MRR@20).}
\label{tab:appendix_window_retailrocket_results_single}
\begin{tabular}{@{}l r r r r r r r r @{}}
\toprule 
Algorithm& MRR@20& HR@20& COV@20& POP@20& MRR@10& HR@10& COV@10& POP@10\\ 
\midrule 
\ssknn & \num{0.332662} & \num{0.580406} & \num{0.272329} & \num{0.051342} & \num{0.328714} & \num{0.524200} & \num{0.170172} & \num{0.059803}\\ 
\knn & \num{0.332349} & \num{0.571038} & \num{0.250499} & \num{0.051928} & \num{0.328752} & \num{0.519516} & \num{0.155456} & \num{0.060582}\\ 
\vsknn & \num{0.326919} & \num{0.581577} & \num{0.272145} & \num{0.061166} & \num{0.322569} & \num{0.519516} & \num{0.172231} & \num{0.068904}\\ 
\sfsknn & \num{0.300770} & \num{0.462920} & \num{0.224390} & \num{0.036432} & \num{0.299278} & \num{0.443794} & \num{0.168481} & \num{0.048572}\\ 
\sr & \num{0.270372} & \num{0.504294} & \num{0.299417} & \num{0.046994} & \num{0.266667} & \num{0.451600} & \num{0.201052} & \num{0.054104}\\ 
\smf & \num{0.269543} & \num{0.556987} & \num{0.312543} & \num{0.056196} & \num{0.264137} & \num{0.478923} & \num{0.198056} & \num{0.061444}\\ 
\ar & \num{0.265118} & \num{0.484778} & \num{0.285639} & \num{0.057562} & \num{0.260942} & \num{0.424668} & \num{0.189311} & \num{0.064483}\\
\bottomrule 
\end{tabular}
\end{table}
\begin{table}[H]
\tiny
\centering
\caption{Hit rate (HR), Mean reciprocal rank (MRR), item coverage (COV), and average popularity (POP) results for a list length of 20, 10, 5, 3, and 1 on the ZALANDO dataset with a single split (sorted by MRR@20).}
\label{tab:appendix_window_zalando_results_single}
\begin{tabular}{@{}l r r r r r r r r @{}}
\toprule 
Algorithm& MRR@20& HR@20& COV@20& POP@20& MRR@10& HR@10& COV@10& POP@10\\ 
\midrule 
\sr & \num{0.305564} & \num{0.497582} & \num{0.524592} & \num{0.059374} & \num{0.303822} & \num{0.472738} & \num{0.380963} & \num{0.060024}\\ 
\mc & \num{0.303644} & \num{0.468563} & \num{0.504156} & \num{0.051610} & \num{0.302474} & \num{0.452150} & \num{0.380856} & \num{0.055026}\\ 
\iknn & \num{0.270904} & \num{0.410313} & \num{0.596650} & \num{0.032728} & \num{0.269353} & \num{0.388271} & \num{0.431925} & \num{0.035827}\\ 
\gru & \num{0.267498} & \num{0.483217} & \num{0.290347} & \num{0.072912} & \num{0.264620} & \num{0.442028} & \num{0.222961} & \num{0.074225}\\ 
\ar & \num{0.264665} & \num{0.483006} & \num{0.430596} & \num{0.072806} & \num{0.262350} & \num{0.449586} & \num{0.311008} & \num{0.072659}\\ 
\smf & \num{0.252770} & \num{0.463330} & \num{0.329088} & \num{0.075164} & \num{0.249838} & \num{0.421505} & \num{0.248270} & \num{0.075975}\\ 
\sfsknn & \num{0.250807} & \num{0.451397} & \num{0.419363} & \num{0.046234} & \num{0.249970} & \num{0.439979} & \num{0.323279} & \num{0.053457}\\
\bottomrule 
\end{tabular}
\end{table}

\sisetup{
	round-mode          = places, 
	round-precision     = 4, 
}

\begin{table}[H]
\tiny
\centering
\caption{Hit rate (HR), Mean reciprocal rank (MRR), item coverage (COV), and average popularity (POP) results for a list length of 20, 10, 5, 3, and 1 on the 8TRACKS dataset with a single split (sorted by MRR@20).}
\label{tab:appendix_window_8tracks_results_single}
\begin{tabular}{@{}l r r r r r r r r @{}}
\toprule 
Algorithm& MRR@20& HR@20& COV@20& POP@20& MRR@10& HR@10& COV@10& POP@10\\ 
\midrule 
\ar & \num{0.013493} & \num{0.040956} & \num{0.776851} & \num{0.039894} & \num{0.012586} & \num{0.027597} & \num{0.577200} & \num{0.054473}\\ 
\sr & \num{0.012282} & \num{0.032873} & \num{0.887539} & \num{0.024268} & \num{0.011629} & \num{0.023316} & \num{0.726141} & \num{0.030246}\\ 
\smf & \num{0.011463} & \num{0.047574} & \num{0.077191} & \num{0.119722} & \num{0.010206} & \num{0.029041} & \num{0.055622} & \num{0.130318}\\ 
\vsknn & \num{0.011012} & \num{0.049017} & \num{0.718030} & \num{0.028991} & \num{0.009794} & \num{0.031256} & \num{0.531670} & \num{0.032244}\\ 
\mc & \num{0.010140} & \num{0.023445} & \num{0.836479} & \num{0.015207} & \num{0.009762} & \num{0.017926} & \num{0.705039} & \num{0.017893}\\ 
\knn & \num{0.009769} & \num{0.043805} & \num{0.612202} & \num{0.027189} & \num{0.008592} & \num{0.026746} & \num{0.434283} & \num{0.029841}\\ 
\ssknn & \num{0.009655} & \num{0.040154} & \num{0.854284} & \num{0.019744} & \num{0.008713} & \num{0.026509} & \num{0.646508} & \num{0.023827}\\
\bottomrule 
\end{tabular}
\end{table}
\begin{table}[H]
\tiny
\centering
\caption{Hit rate (HR), Mean reciprocal rank (MRR), item coverage (COV), and average popularity (POP) results for a list length of 20, 10, 5, 3, and 1 on the AOTM dataset with a single split (sorted by MRR@20).}
\label{tab:appendix_window_aotm_results_single}
\begin{tabular}{@{}l r r r r r r r r @{}}
\toprule 
Algorithm& MRR@20& HR@20& COV@20& POP@20& MRR@10& HR@10& COV@10& POP@10\\ 
\midrule 
\sfsknn & \num{0.027534} & \num{0.043958} & \num{0.859142} & \num{0.082830} & \num{0.027275} & \num{0.040258} & \num{0.767294} & \num{0.102502}\\ 
\smf & \num{0.020407} & \num{0.046843} & \num{0.926151} & \num{0.094061} & \num{0.019671} & \num{0.036083} & \num{0.829373} & \num{0.094128}\\ 
\gru & \num{0.015365} & \num{0.042702} & \num{0.652258} & \num{0.166456} & \num{0.014569} & \num{0.031229} & \num{0.508119} & \num{0.175874}\\ 
\sr & \num{0.015229} & \num{0.044908} & \num{0.943946} & \num{0.106092} & \num{0.014327} & \num{0.031772} & \num{0.813950} & \num{0.114280}\\ 
\mc & \num{0.013391} & \num{0.034827} & \num{0.899626} & \num{0.081337} & \num{0.012784} & \num{0.026239} & \num{0.788889} & \num{0.090178}\\ 
\ar & \num{0.011878} & \num{0.042634} & \num{0.852339} & \num{0.140925} & \num{0.010901} & \num{0.028276} & \num{0.672314} & \num{0.153601}\\ 
\vsknn & \num{0.010375} & \num{0.072098} & \num{0.797101} & \num{0.156728} & \num{0.008258} & \num{0.041548} & \num{0.604879} & \num{0.166183}\\
\bottomrule 
\end{tabular}
\end{table}
\begin{table}[H]
\tiny
\centering
\caption{Hit rate (HR), Mean reciprocal rank (MRR), item coverage (COV), and average popularity (POP) results for a list length of 20, 10, 5, 3, and 1 on the 30MUSIC dataset with a single split (sorted by MRR@20).}
\label{tab:appendix_window_30music_results_single}
\begin{tabular}{@{}l r r r r r r r r @{}}
\toprule 
Algorithm& MRR@20& HR@20& COV@20& POP@20& MRR@10& HR@10& COV@10& POP@10\\ 
\midrule 
\sr & \num{0.268957} & \num{0.374406} & \num{0.590405} & \num{0.037331} & \num{0.267206} & \num{0.349912} & \num{0.461900} & \num{0.038900}\\ 
\mc & \num{0.265287} & \num{0.330150} & \num{0.400058} & \num{0.028275} & \num{0.264576} & \num{0.320274} & \num{0.345480} & \num{0.033950}\\ 
\gru & \num{0.235387} & \num{0.365110} & \num{0.402885} & \num{0.066538} & \num{0.233416} & \num{0.337169} & \num{0.307656} & \num{0.066858}\\ 
\smf & \num{0.214500} & \num{0.361394} & \num{0.397449} & \num{0.060805} & \num{0.212105} & \num{0.327477} & \num{0.301271} & \num{0.060453}\\ 
\sfsknn & \num{0.212251} & \num{0.332008} & \num{0.340356} & \num{0.028437} & \num{0.211780} & \num{0.325805} & \num{0.292922} & \num{0.036636}\\ 
\iknn & \num{0.135228} & \num{0.341170} & \num{0.675810} & \num{0.021873} & \num{0.131857} & \num{0.294275} & \num{0.504323} & \num{0.023407}\\ 
\vsknn & \num{0.119232} & \num{0.413400} & \num{0.438381} & \num{0.058878} & \num{0.112988} & \num{0.326250} & \num{0.291052} & \num{0.059862}\\
\bottomrule 
\end{tabular}
\end{table}
\begin{table}[H]
\tiny
\centering
\caption{Hit rate (HR), Mean reciprocal rank (MRR), item coverage (COV), and average popularity (POP) results for a list length of 20, 10, 5, 3, and 1 on the NOWPLAYING dataset with a single split (sorted by MRR@20).}
\label{tab:appendix_window_nowplaying_results_single}
\begin{tabular}{@{}l r r r r r r r r @{}}
\toprule 
Algorithm& MRR@20& HR@20& COV@20& POP@20& MRR@10& HR@10& COV@10& POP@10\\ 
\midrule 
\sr & \num{0.085584} & \num{0.182540} & \num{0.462927} & \num{0.030854} & \num{0.083129} & \num{0.146639} & \num{0.339042} & \num{0.034579}\\ 
\mc & \num{0.081283} & \num{0.147432} & \num{0.357601} & \num{0.024794} & \num{0.079754} & \num{0.125537} & \num{0.283274} & \num{0.028749}\\ 
\sfsknn & \num{0.078650} & \num{0.160177} & \num{0.301525} & \num{0.026379} & \num{0.077422} & \num{0.143091} & \num{0.238317} & \num{0.033312}\\ 
\smf & \num{0.078158} & \num{0.188095} & \num{0.355993} & \num{0.032182} & \num{0.075250} & \num{0.145378} & \num{0.258505} & \num{0.033217}\\ 
\gru & \num{0.077068} & \num{0.179225} & \num{0.220233} & \num{0.052270} & \num{0.074155} & \num{0.136975} & \num{0.164661} & \num{0.056787}\\ 
\vsknn & \num{0.067028} & \num{0.229132} & \num{0.335846} & \num{0.044516} & \num{0.062427} & \num{0.162745} & \num{0.224772} & \num{0.049689}\\ 
\ssknn & \num{0.066934} & \num{0.240616} & \num{0.343615} & \num{0.040740} & \num{0.061827} & \num{0.167414} & \num{0.218576} & \num{0.046819}\\
\bottomrule 
\end{tabular}
\end{table}

\sisetup{
	round-mode          = places, 
	round-precision     = 3, 
}

\begin{table}[H]
\tiny
\centering
\caption{Hit rate (HR), Mean reciprocal rank (MRR), item coverage (COV), and average popularity (POP) results for a list length of 20, 10, 5, 3, and 1 on the CLEF dataset with a single split (sorted by MRR@20).}
\label{tab:appendix_window_clef_results_single}
\begin{tabular}{@{}l r r r r r r r r @{}}
\toprule 
Algorithm& MRR@20& HR@20& COV@20& POP@20& MRR@10& HR@10& COV@10& POP@10\\ 
\midrule 
\gru & \num{0.253342} & \num{0.724281} & \num{0.153846} & \num{0.050570} & \num{0.242340} & \num{0.564453} & \num{0.123077} & \num{0.063601}\\ 
\sr & \num{0.213436} & \num{0.622531} & \num{0.753846} & \num{0.055094} & \num{0.201593} & \num{0.450697} & \num{0.715385} & \num{0.083689}\\ 
\mc & \num{0.213318} & \num{0.624562} & \num{0.754545} & \num{0.056264} & \num{0.201758} & \num{0.459113} & \num{0.723776} & \num{0.080135}\\ 
\smf & \num{0.206554} & \num{0.646295} & \num{0.765035} & \num{0.041706} & \num{0.194426} & \num{0.471786} & \num{0.709091} & \num{0.041691}\\ 
\vsknn & \num{0.196587} & \num{0.683345} & \num{0.755944} & \num{0.053321} & \num{0.182975} & \num{0.487209} & \num{0.681818} & \num{0.079510}\\ 
\ssknn & \num{0.189707} & \num{0.669952} & \num{0.764336} & \num{0.051940} & \num{0.175398} & \num{0.462999} & \num{0.681818} & \num{0.081052}\\ 
\knn & \num{0.186094} & \num{0.660559} & \num{0.760140} & \num{0.052229} & \num{0.171771} & \num{0.452780} & \num{0.674825} & \num{0.082596}\\
\bottomrule 
\end{tabular}
\end{table}

\end{document}